\documentclass[12pt]{article}
\pdfoutput=1
\usepackage[utf8]{inputenc}
\usepackage{epsf,amsmath,amssymb,graphicx,dcolumn}
\usepackage{caption}
\usepackage{subcaption}
\usepackage{scalefnt,ulem,pstricks}
\usepackage{booktabs,multirow,tabularx}
\usepackage{cite,hyperref}
\usepackage{cleveref}
\usepackage{color}
\usepackage{rotating}
\usepackage{microtype}
\definecolor{lightblue}{rgb}{.7,.8,1}
\def\refse#1{\mbox{Section~\ref{#1}}}
\renewcommand{\thefootnote}{\fnsymbol{footnote}}
\def\tL{{\tilde L}}
\newcommand\nn{\nonumber}
\newcommand\f[2]{\frac{#1}{#2}} 
\newcommand{\abbrev}{\scalefont{1.}}

\def\bal#1\eal{\begin{align}#1\end{align}}

\newcommand{\as}{\alpha_{\mathrm{S}}}

\newcommand{\openloops}{{\sc OpenLoops}}
\newcommand{\munich}{{\sc Munich}}
\newcommand{\Collier}{{\sc Collier}}
\newcommand{\CutTools}{{\sc CutTools}}
\newcommand{\OneLOop}{{\sc OneLOop}}

\newcommand{\mcfm}{{\abbrev MCFM}}

\newcommand{\lhc}{{\abbrev LHC}}

\newcommand{\bsm}{{\abbrev BSM}}

\newcommand{\powheg}{{\abbrev POWHEG}}
\newcommand{\pythia}[1]{{\sc Pythia{#1}}}

\newcommand{\atl}{{\abbrev ATLAS}}
\newcommand{\cms}{{\abbrev CMS}}

\newcommand{\zz}{\ensuremath{ZZ}}
\newcommand{\ww}{\ensuremath{W^+W^-}}

\newcommand{\vvp}{\ensuremath{VV^\prime}}

\newcommand{\pt}{\ensuremath{p_T}}

\newcommand{\llog}{\text{\abbrev LL}}
\newcommand{\nll}{\text{\abbrev NLL}}
\newcommand{\nnll}{\text{\abbrev NNLL}}
\newcommand{\lo}{\text{\abbrev LO}}
\newcommand{\nlo}{\text{\abbrev NLO}}
\newcommand{\nnlo}{\text{\abbrev NNLO}}

\newcommand{\sm}{{\abbrev SM}}
\newcommand{\ew}{{\abbrev EW}}
\newcommand{\qcd}{{\abbrev QCD}}

\newcommand{\fs}[1]{#1{\abbrev FS}}

\newcommand{\plus}{{\abbrev +}}
\newcommand{\citere}[1]{Ref.\cite{#1}}
\newcommand{\citeres}[1]{Refs.\cite{#1}}
\newcommand{\eqn}[1]{Eq.\,(\ref{#1})}

\newcommand{\fig}[1]{Fig.\,\ref{#1}}
\newcommand{\figs}[1]{Figs.\,\ref{#1}}
\newcommand{\tab}[1]{Tab.\,\ref{#1}}
\newcommand{\sct}[1]{Section~\ref{#1}}

\newcommand{\muF}{\mu_{F}}
\newcommand{\muR}{\mu_{R}}
\newcommand{\Qres}{Q}

\newcommand{\msbar}{\overline{\text{MS}}}

\newcommand\Tstrut{\rule{0pt}{3.0ex}}         
\newcommand\Bstrut{\rule[-1.5ex]{0pt}{0pt}}   

\Crefname{figure}{Fig.}{Figs.}
\begin{document}
\begin{titlepage}
\renewcommand{\thefootnote}{\fnsymbol{footnote}}
\begin{flushright}
ZU-TH 20/15\\
MITP/15-051
\end{flushright}
\vspace*{2cm}

\begin{center}
{\Large \bf Transverse-momentum resummation\\[0.4cm] 
for vector-boson pair production at \nnll{}\plus{}\nnlo{}}
\end{center}

\par \vspace{2mm}
\begin{center}
{\bf Massimiliano Grazzini$^{(a)}$\footnote{On leave of absence from INFN, Sezione di Firenze, Sesto Fiorentino, Florence, Italy.}},
{\bf Stefan Kallweit$^{(b)}$},\\[0.2cm]
{\bf Dirk Rathlev$^{(a)}$} and {\bf Marius Wiesemann$^{(a)}$}

\vspace{5mm}

$^{(a)}$ Physik-Institut, Universit\"at Z\"urich, 
CH-8057 Z\"urich, Switzerland 

$^{(b)}$ PRISMA Cluster of Excellence, Institute of Physics,\\[0.1cm]
Johannes Gutenberg University, D-55099 Mainz, Germany

\end{center}

\par \vspace{2mm}
\begin{center} {\large \bf Abstract} \end{center}
\begin{quote}
\pretolerance 10000

We consider the transverse-momentum (\pt{}) distribution of \zz{} and \ww{} boson pairs produced in hadron collisions. At small \pt{}, the logarithmically enhanced contributions due to multiple soft-gluon emission are resummed to all orders in QCD perturbation theory. At intermediate and large values of \pt{}, we consistently combine resummation with the known fixed-order results. We exploit the most advanced perturbative information that is available at present: next-to-next-to-leading logarithmic resummation combined with the next-to-next-to-leading fixed-order calculation. 
After integration over \pt{}, we recover the known next-to-next-to-leading order result for the inclusive cross section. We present numerical results at the LHC, together with an estimate of the corresponding uncertainties. We also study the rapidity dependence of the \pt{} spectrum and
we consider \pt{} efficiencies at different orders of resummed and fixed-order perturbation theory.
    
\end{quote}

\vspace*{\fill}
\begin{flushleft}
July 2015

\end{flushleft}
\end{titlepage}

\section{Introduction}
\label{sec:intro}

Run 1 of the Large Hadron Collider (\lhc) has been a great success for the Standard Model (\sm). The collected data are in good agreement with the theoretical predictions so far and led to the discovery \cite{Aad:2012tfa,Chatrchyan:2012ufa} of a resonance at a mass of $125$\,GeV, which appears to be fully consistent with the \sm{} Higgs boson.
Among the most important reactions at hadron colliders is the production of vector-boson pairs.
This class of processes gives access to the vector-boson trilinear couplings which may be modified
in a large set of Beyond the Standard Model (\bsm{}) theories. Even small deviations in both the production rate and the shape
of distributions could be a signal of new physics. Anomalous couplings related to
vector-boson pair production have been constrained first by {\abbrev LEP2}, and later by the Tevatron
for larger invariant masses. \atl{} and \cms{} will continue to tighten the bounds on anomalous couplings,
especially with increasing sensitivity during Run 2 of the \lhc{}.\footnote{See \citere{Wang:2014uea} and references therein.} 

On the other hand, vector-boson pair production constitutes an irreducible background
to new-physics searches as well as Higgs studies. Particularly important
are the off-shell effects below
the \zz{} and \ww{} thresholds, relevant for the Higgs signal region,
and the high-mass tail used
to extract the width of the Higgs boson \cite{Kauer:2012hd,Caola:2013yja,Campbell:2013wga}.
Furthermore, Higgs boson measurements, in particular in the \ww{} channel, strongly rely on the background rejection through specific categories based on the transverse momenta of final-state particles, such as the classification into jet bins or in the Higgs transverse momentum and related variables.
An accurate modelling of the respective observables for both signal and backgrounds
is crucial for such analyses.

The first precise predictions for \zz{} production at hadron colliders in the \sm{} were obtained at the next-to-leading order (\nlo{}) already more than 20 years ago for stable $Z$ bosons \cite{Ohnemus:1990za,Mele:1990bq}, and the leptonic decays were added in \citere{Ohnemus:1994ff}. The full spin correlations and off-shell effects at the \nlo{} were first included in \citeres{Campbell:1999ah,Dixon:1999di} using the corresponding one-loop helicity amplitudes \cite{Dixon:1998py}. An important loop-induced contribution proceeds through gluon fusion; it is enhanced by the gluon densities and was first computed for on-shell $Z$ bosons in \citeres{Dicus:1987dj,vanderBij:1988fb}, while leptonic decays were included later \cite{Matsuura:1991pj,Zecher:1994kb,Binoth:2008pr}. All the contributions to \zz{} production discussed so far are implemented in the numerical program \mcfm{} \cite{Campbell:2011bn}. Electroweak (\ew) corrections were evaluated in \citere{Bierweiler:2013dja,Baglio:2013toa}, while on-shell $ZZ+$jet production is known through \nlo{} \qcd{} \cite{Binoth:2009wk,Binoth:2010nha}. 
Recently, the inclusive cross section for the production of \zz{} at the next-to-next-to-leading order (\nnlo{}) has been presented \cite{Cascioli:2014yka}. 

The production of \ww{} pairs constitutes the largest cross section among all massive vector-boson pair production modes. On the other hand, its leptonic decay ($\ww \to l^+l^-\nu\bar\nu$) embodies the most challenging experimental signature since the presence of two neutrinos prohibits a reconstruction of mass peaks. Therefore, a precise understanding of both the signal and the background is required.

The \ww{} cross sections for on-shell $W$ bosons at the \nlo{} \cite{Ohnemus:1991kk,Frixione:1993yp} as well as the gluon-fusion component \cite{Dicus:1987dj,Glover:1988fe} have been known for decades. Also in this case, spin and off-shell effects were included in the \nlo{} prediction \cite{Campbell:1999ah,Dixon:1999di} after the relevant one-loop helicity amplitudes had been computed \cite{Dixon:1998py}.
Leptonic decays for the loop-induced gluon-fusion contribution were considered in \citeres{Binoth:2005ua,Binoth:2006mf}. More recently, also interference effects with the Higgs production mode through gluon fusion were determined \cite{Campbell:2011cu}. Analogously to \zz{} production, all the contributions to \ww{} production discussed
so far are implemented in \mcfm\,\cite{Campbell:2011bn}.
Furthermore, \ew{} corrections have been evaluated \cite{Bierweiler:2012kw,Baglio:2013toa,Billoni:2013aba}. On-shell $\ww$ production in association with one jet has been studied through \nlo{} \qcd{} in \citeres{Dittmaier:2007th,Campbell:2007ev,Dittmaier:2009un}. 
Detailed Monte Carlo simulations of $e^+\nu_e\mu^-\bar\nu_\mu$ production in association with up to 
one jet at \nlo{} have been presented in Ref.~\cite{Cascioli:2013gfa}.
Recently, the first \nnlo{} results for the inclusive \ww{} cross section have been obtained \cite{Gehrmann:2014fva}.

Due to the very recent computation of the $q\bar{q}\to VV^\prime$ helicity amplitudes at two-loop order \cite{Caola:2014iua,Gehrmann:2015ora}, the inclusion of the off-shell effects and the leptonic decays in the \nnlo{} cross section is expected in the near future.
In the meanwhile, also the calculation of $gg\to VV^\prime$ helicity amplitudes has been performed at two-loop order \cite{Caola:2015ila,vonManteuffel:2015msa}. 
This renders the evaluation of \nlo{} \qcd{} corrections to the gluon-fusion channel feasible.

The production of \zz{} pairs in hadron collisions has been measured extensively at the Tevatron and the \lhc{} (see \citeres{CDF:2011ab,Abazov:2012cj,Aad:2012awa,Chatrchyan:2012sga,ATLAS:2013gma,Chatrchyan:2013oev,CMS:2014xja} for some recent results). 
The \ww{} cross section has also been measured already at the Tevatron, see e.g. \citere{Abazov:2011cb}, and at the \lhc{} both at 7 TeV \cite{ATLAS:2012mec,Chatrchyan:2013yaa} and 8 TeV \cite{Chatrchyan:2013oev,ATLAS-CONF-2014-033,CMS:2015uda}.
The \atl{} collaboration recently reported an excess \cite{ATLAS-CONF-2014-033} with respect to the \sm{} prediction,
which has drawn a lot of attention on the \ww{} process, 
since the \ww{} final state is a typical signature in many \bsm{} scenarios \cite{Morrissey:2009tf}. In the meanwhile, the excess has been alleviated to a significant degree by the recent \nnlo{} computation \cite{Gehrmann:2014fva}.
The more recent measurement by the \cms{} collaboration \cite{CMS:2015uda} is in good agreement with the \nnlo{} prediction.

The transverse-momentum distributions of \zz{} and \ww{} pairs are among the 
most important differential observables for these processes. The \pt{}
spectrum has already been measured in the case of \zz{} production \cite{CMS:2014xja} at the \lhc{}.
Transverse-momentum resummation for  \zz{} and \ww{} production
has been studied in \citeres{Balazs:1998bm,Grazzini:2005vw,Frederix:2008vb,Wang:2013qua,Meade:2014fca}.
In all these calculations the resummed computation is essentially performed up to next-to-leading logarithmic (\nll{}) accuracy
(next-to-next-to-leading logarithmic (\nnll) effects in the Sudakov exponent are considered in \citeres{Grazzini:2005vw,Wang:2013qua,Meade:2014fca})
and matched to the fixed-order result up to the first order
in the \qcd{} coupling $\as$.

In this paper we consider transverse-momentum resummation for the production of both 
\zz{}  and \ww{} pairs.
We use the formalism of \citere{Bozzi:2005wk} to perform the first computation of the \pt{} spectrum
at full \nnll{} accuracy matched to the ${\cal O}(\as^2)$ fixed-order result valid at large \pt{}.
Although our focus is on the 
inclusive \pt{} spectrum of the \zz{} and \ww{} system, our computation is fully 
differential in the degrees of freedom of the vector bosons and allows us to include 
their \ew{} decays, once the helicity amplitudes are implemented\footnote{The analogous computations for Higgs, single vector-boson production, and diphoton production are presented
in \citeres{deFlorian:2012mx}, \cite{Catani:2015vma} and \cite{Cieri:2015rqa}, respectively.}.
The \pt{}-resummation formalism of \citere{Bozzi:2005wk} is closely related to the subtraction method 
of \citere{Catani:2007vq}, which was
used to compute the \nnlo{} cross section for these processes \cite{Cascioli:2014yka,Gehrmann:2014fva}. 
For \ww{} production we employ 
the four-flavour scheme (\fs{4}) and remove all contributions with final-state bottom quarks from our 
computation of the \ww{} transverse-momentum distribution in order to eliminate the 
contamination from $t\bar{t}$ and $Wt$ production. The difference to the prediction in the
five-flavour scheme (\fs{5}),
where such terms have to be consistently subtracted, has been shown to 
be small for the \nnlo{} inclusive cross section \cite{Gehrmann:2014fva}. Furthermore, we neglect the
loop-induced gluon-fusion contribution throughout this paper, since, up to \nnlo{},
it contributes only at $\pt=0$.

We note that
in the CMS measurement reported in \citere{CMS:2015uda} an approximate \nnll{} prediction \cite{Meade:2014fca} of the \pt{}
spectrum has been used to correct the spectrum from the Monte Carlo simulation.
Along these lines,
the computation reported in this paper will be useful
in order to validate the predictions obtained from Monte Carlo simulations for both \zz{} and \ww{} production,
as done in the case of Higgs boson production with the calculation of \citere{Bozzi:2005wk}.

The manuscript is organized as follows. In Section \ref{sec:resum}
we review the transverse-momentum resummation formalism applied
to vector-boson pair production. In Section \ref{sec:results} we report our numerical results, starting with remarks on the choice of the resummation scale in \sct{sec:resscale}. In \refse{sec:inclpt} we present our numerical predictions for the inclusive \pt{} spectrum and study the ensuing uncertainties. In \refse{sec:rap} we analyse the behaviour of the spectrum at different rapidities of the vector-boson pair. In \refse{sec:pteff} we investigate \pt{} efficiencies at different orders in resummed and fixed-order perturbation theory. In \refse{sec:summa} we summarize our results.

\section{Transverse-momentum resummation for vector-boson\\ pair production}
\label{sec:resum}

In this Section we recall the main points of the transverse-momentum resummation formalism we use in this paper.
For a more detailed discussion the reader is referred to Refs.~\cite{Bozzi:2005wk,Bozzi:2007pn,Catani:2013tia}.

We consider the inclusive hard-scattering process
\begin{equation}
h_1(P_1)+h_2(P_2)\to V(p_3)+V^\prime(p_4)+X\,,
\end{equation}
where the collision of the two hadrons $h_1$ and $h_2$ with momenta $P_1$ and
$P_2$ produces the two vector bosons of momenta $p_3$ and $p_4$.
In the center-of-mass frame the momentum
of the vector-boson pair $q=p_3+p_4$ is fully specified by
the invariant mass $M^2=(p_3+p_4)^2$, the rapidity $y=\f{1}{2}\ln\f{q\cdot P_1}{q\cdot P_2}$, and the transverse-momentum vector ${\boldsymbol \pt}$.

The kinematics of the vector bosons
is fully determined by the vector-boson pair momentum $q^\mu=p^\mu_3+p^\mu_4$ (with $p_3^2=m_V^2$ and $p_4^2=m_{V'}^2$), and by the additional and independent variables that
specify the angular distribution of the vector bosons
with respect to $q^\mu$.  Throughout this paper we always
consider quantities
which are inclusive over these angular variables.

According to the \qcd{} factorization theorem, the differential cross section can be written as
\begin{align}
\label{dcross}
\f{d\sigma^{VV^\prime}}{dM^2\,d p_T^2dy}(y,p_T,M,s)&= \sum_{a_1,a_2}
\int_0^1 dx_1 \,\int_0^1 dx_2 \,f_{a_1/h_1}(x_1,\mu_F^2)
\,f_{a_2/h_2}(x_2,\mu_F^2)\nn\\
&\times\f{d{\hat \sigma}^{VV^\prime}_{a_1a_2}}{dM^2 d p_T^2\,d{\hat y}}({\hat y},p_T,M,{\hat s},
\as(\mu_R^2),\mu_R^2,\mu_F^2) 
\,,
\end{align}
where $f_{a/h}(x,\mu_F^2)$ ($a=q,{\bar q},g$) are the density functions of parton $a$ in hadron $h$ at the
factorization scale $\muF$; $\muR$ is the renormalization scale\footnote{Throughout the paper we use parton densities in the $\msbar$ factorization scheme and $\as(q^2)$ is the \qcd{} running coupling in the $\msbar$ renormalization scheme.};
$d{\hat \sigma}^{VV^\prime}_{a_1a_2}$ is the partonic cross section.
The rapidity ${\hat y}$ and the center-of-mass energy ${\hat s}$ of the partonic scattering process
are related to the corresponding hadronic variables $y$ and $s$ by
\begin{equation}
{\hat y}=y-\f{1}{2}\ln\f{x_1}{x_2},~~~~~~~~~{\hat s}=x_1x_2 s\, .
\end{equation}
When the transverse momentum $\pt$ of the vector-boson pair is of the same order as the invariant mass $M$,
the \qcd{} perturbative expansion is controlled by a single expansion parameter, $\as(M)$, and fixed-order
calculations can be safely applied. In this region, \qcd{} radiative corrections are known to ${\cal O}(\as^2)$ \cite{Binoth:2009wk,Binoth:2010nha,Dittmaier:2007th,Campbell:2007ev,Dittmaier:2009un}.
When $\pt\ll M$ the convergence of the perturbative expansion is spoiled by the presence of large logarithmic terms
of the form $\as^n\ln^m(M^2/\pt^2)$, that need to be resummed to all orders.

The resummation is performed at the level of the partonic cross section,
which is decomposed as
\begin{equation}
\label{resplusfin}
\f{d{\hat \sigma}^{VV^\prime}_{a_1a_2}}{dM^2dp_T^2d{\hat y}}=
\f{d{\hat \sigma}^{VV^\prime,\rm (res.)}_{a_1a_2}}{dM^2dp_T^2d{\hat y}}+
\f{d{\hat \sigma}^{VV^\prime,\rm (fin.)}_{a_1a_2}}{dM^2dp_T^2d{\hat y}}\, .
\end{equation}
The first term
on the right-hand side of \eqn{resplusfin}
contains all the logarithmically-enhanced contributions at small \pt{}
and has to be evaluated by resumming them to all orders.
The second term is instead free of such contributions and
can thus be evaluated at fixed order in perturbation theory. 

Resummation is based on the factorization of soft and collinear radiation and is viable in the impact-parameter ($b$) space, where the kinematical constraint of momentum conservation and the factorization of the phase space can be consistently taken into account \cite{Parisi:1979se,Curci:1979bg,Collins:1984kg}.
Using the Bessel transformation between the conjugate variables 
$\pt$ and $b$,
the resummed component is expressed as \cite{Collins:1984kg,Bozzi:2005wk}
\begin{equation}
\label{resum}
\f{d{\hat \sigma}_{a_1a_2}^{VV^\prime,(\rm res.)}}{dM^2dp_T^2\,d{\hat y}}=
\f{M^2}{\hat s} \;
\int_0^\infty db \; \frac{b}{2} \;J_0(b p_T) 
\;{\cal W}^{\;VV^\prime}_{a_1a_2}(b,{\hat y},M,\hat s;\as,\mu_R^2,\mu_F^2) \;,
\end{equation}
where $J_0(x)$ is the $0$-order Bessel function. In the case of fully inclusive 
\pt{} resummation, the rapidity dependence is integrated out in Eq.~(\ref{resum}). In that case it is convenient to consider Mellin moments with respect to the variable $z = M^2/\hat s$. However, in order to retain the rapidity dependence in the resummed cross section we take the `double' $(N_1,N_2)$ Mellin moments
with respect to the variables 
$z_1=e^{+\hat y}M/{\sqrt{\hat s}}$ and $z_2=e^{-\hat y}M/{\sqrt{\hat s}}$ at fixed $M$,
\begin{equation}
{\cal W}^{VV^\prime}_{(N_1,N_2)}(b,M;\as,\muR^2,\muF^2)
=\int_0^1dz_1\, z_1^{N_1-1}\int_0^1dz_2\, z_2^{N_2-1}
{\cal W}^{VV^\prime}(b,{\hat y},M,\hat s;\as,\muR^2,\muF^2)\, ,
\end{equation}
and organize the structure of ${\cal W}_{VV^\prime}$ in the following exponential form \cite{Bozzi:2007pn},
\begin{align}
\label{wtilde}
{\cal W}^{VV^\prime}_{(N_1,N_2)}(b,M;\as,\muR^2,\muF^2)
&={\cal H}^{VV^\prime}_{(N_1,N_2)}\left(M; 
\as,M^2/\mu_R^2,M^2/\mu_F^2,M^2/Q^2 \right) \nonumber \\
&\times \exp\{{\cal G}_{(N_1,N_2)}(\as,L,M^2/\muR^2,M^2/Q^2)\}\, ,
\end{align}
where we have defined the logarithmic expansion parameter $L$ as
\begin{equation}
\label{logpar}
L=\ln\f{Q^2b^2}{b_0^2}\, ,
\end{equation}
and $b_0=2e^{-\gamma_E}$ ($\gamma_E=0.5772...$ is the Euler number).
The scale $Q$ appearing in Eqs.~(\ref{wtilde},~\ref{logpar}),
named resummation scale in Ref.~\cite{Bozzi:2005wk},
parametrizes the
arbitrariness in the resummation procedure, 
and has to be chosen of the order of the hard scale $M$.
Variations of $Q$ around its reference value can be exploited to estimate the size of yet uncalculated higher-order logarithmic contributions. Therefore, $Q$ plays a role very 
similar to $\mu_F$ and $\mu_R$ for missing perturbative terms.
The function ${\cal H}^{VV^\prime}_{(N_1,N_2)}$ does not depend on the impact parameter $b$, and therefore includes all the perturbative
terms that behave as constants as $b\to\infty$. It can thus be expanded in powers of $\as=\as(\mu_R^2)$:
\begin{align}
\label{hexpan}
{\cal H}^{VV^\prime}_{(N_1,N_2)}(M,\as;M^2/\mu^2_R,M^2/\mu^2_F,M^2/Q^2)&=
\sigma^{VV^\prime,(0)}(\as,M)\nn\\
&\times\Bigr[ 1+\f{\as}{\pi} \,{\cal H}^{VV^\prime,(1)}_{(N_1,N_2)}(M^2/\mu^2_R,M^2/\mu^2_F,M^2/Q^2) 
\Bigr. \\
&+ \Bigl.
\left(\f{\as}{\pi}\right)^2 
\,{\cal H}^{VV^\prime,(2)}_{(N_1,N_2)}(M^2/\mu^2_R,M^2/\mu^2_F,M^2/Q^2)+\dots \Bigr]\,,\nn 
\end{align}
where $\sigma^{VV^\prime,(0)}$ is the partonic leading-order (\lo{}) cross section.
The exponent ${\cal G}_{(N_1,N_2)}$ includes the complete dependence on $b$ and, in particular, it contains all
the terms that order-by-order in $\as$ are logarithmically divergent as $b\to\infty$.
The logarithmic expansion of ${\cal G}_N$ reads
\begin{align}
\label{exponent}
{\cal G}_{(N_1,N_2)}(\as(\mu^2_R),L;M^2/\mu^2_R,M^2/Q^2)&=L g^{(1)}(\as L)+g^{(2)}_{(N_1,N_2)}(\as L;M^2/\mu_R^2,M^2/Q^2)\nn\\
&+\f{\as}{\pi} g^{(3)}_{(N_1,N_2)}(\as L,M^2/\mu_R^2,M^2/Q^2)+\dots\,,
\end{align}
 where the term $L\, g^{(1)}$ collects the \llog{} contributions, the function $g^{(2)}_{(N_1,N_2)}$ includes
the \nll{} contributions, $g^{(3)}_{(N_1,N_2)}$ controls the \nnll{} terms and so forth.

The resummation of the large logarithmic terms carried out in \eqn{wtilde}, after transforming back to \pt{} space, allows us to obtain
a well behaved transverse-momentum spectrum as $\pt\to 0$. However, the logarithmic expansion parameter $L$ in \eqn{logpar} is divergent as $b\to 0$. This implies that the resummation produces higher-order contributions also in the high-$\pt$ region, which is conjugated to $b\to 0$ after Fourier transformation. In this region the fixed-order cross section is perfectly viable and any resummation effect is necessarily artificial.
To reduce the impact of such contributions,
the logarithmic variable $L$ is replaced by \cite{Bozzi:2005wk}
\begin{equation}
\label{ltilde}
L\to\tL,\quad\quad\quad \tL\equiv \ln \left(\f{Q^2 b^2}{b_0^2}+1\right)\, .
\end{equation}
The variables $L$ and $\tL$ are equivalent when $Qb\gg 1$ but they have a very different behaviour as $b\to 0$. When $Qb\ll 1$, $\tL\to 0$ and ${\cal G}_{(N_1,N_2)}\to 1$.
Moreover, since the behaviour of the Sudakov form factor at $b=0$ is related
to the integral over $\pt$,
the replacement in \eqn{ltilde} allows us to
enforce a {\it unitarity constraint}
such that the fixed-order prediction is recovered upon integration over $\pt$.

A well known property of the formalism of Ref.~\cite{Bozzi:2005wk} is that
the process dependence (as well as the factorization scale and scheme dependence) is fully encoded
in the hard function ${\cal H}_{VV^\prime}$.
In other words, the functions $g^{(i)}$ are universal: they depend only on the channel
in which the process occurs at Born level ($q{\bar q}$ annihilation
in the case of vector-boson pair production). Their explicit expressions up to $i=3$
are given in Ref.~\cite{Bozzi:2005wk}
in terms of the universal perturbative coefficients
$A_q^{(1)}$, $A_q^{(2)}$, $A_q^{(3)}$,
${\tilde B}_{q,N}^{(1)}$, ${\tilde B}_{q,N}^{(2)}$.
In particular, the \llog{} function $g^{(1)}$ depends on the coefficient $A_q^{(1)}$, the \nll{} function $g_{(N_1,N_2)}^{(2)}$
also depends on $A_q^{(2)}$ and ${\tilde B}_q^{(1)}$ \cite{Kodaira:1981nh} and the \nnll{} function $g_{(N_1,N_2)}^{(3)}$ also depends on $A_q^{(3)}$ \cite{Becher:2010tm} and ${\tilde B}_{q,N}^{(2)}$ \cite{Davies:1984hs,deFlorian:2000pr,deFlorian:2001zd}.

The hard coefficients ${\cal H}^{VV^\prime}$ depend on the process we want to consider. The first order coefficients ${\cal H}^{VV^\prime,(1)}$ are known since long time \cite{deFlorian:2000pr,deFlorian:2001zd}: 
they can be obtained from the one-loop scattering amplitudes $q{\bar q}\to VV^\prime$ by using a process independent relation.
By exploiting the expressions of ${\cal H}^{(2)}$ for single Higgs \cite{Catani:2011kr} and vector-boson \cite{Catani:2012qa} production,
in \citere{Catani:2013tia} such a relation has been extended to ${\cal O}(\as^2)$: this implies that
the two-loop amplitude for $q{\bar q}\to VV^\prime$ 
is the only process-dependent information needed to obtain the coefficient ${\cal H}^{VV^\prime,(2)}$.

We now turn to
the finite component of the transverse-momentum spectrum, i.e.\ the second term on the right hand side of \eqn{resplusfin}.
Since $d{\hat \sigma}^{VV^\prime,\rm (fin.)}_{ab}$ does not contain large logarithmic terms
in the small-$\pt$ region,
it can be evaluated by truncating the perturbative series
at a given fixed order.
In practice, the finite component is computed starting from the customary
fixed-order (f.o.) perturbative truncation of the partonic cross section and
subtracting the expansion of the resummed cross section in \eqn{resum} at the same perturbative order:
\begin{equation}
\label{resfin}
\Biggl[\f{d{\hat \sigma}^{VV^\prime,\rm (fin.)}_{a_1a_2}}{dM^2dp_T^2d{\hat y}}\Biggr]_{\rm f.o.}
=\Biggl[\f{d{\hat \sigma}^{VV^\prime}_{a_1a_2}}{dM^2dp_T^2d{\hat y}}\Biggr]_{\rm f.o.}-
\Biggl[\f{d{\hat \sigma}^{VV^\prime, \rm (res.)}_{a_1a_2}}{dM^2dp_T^2d{\hat y}}\Biggr]_{\rm f.o.}\,.
\end{equation} 
At least formally, this matching procedure 
between resummed and finite contributions guarantees
to achieve a uniform theoretical accuracy 
over the entire range of transverse momenta.
At large values of \pt{}, the resummation procedure cannot improve the fixed-order result,
and the resummation (and matching) procedure is eventually superseded by the
customary fixed-order calculations.

In summary,
the inclusion of the functions $g^{(1)}$, $g^{(2)}_{(N_1,N_2)}$,
${\cal H}^{VV^\prime,(1)}$ in the resummed component,
together with the evaluation of the finite component at \nlo{} (i.e.\ ${\cal O}(\as)$),
allows us to perform the resummation at \nll{}\plus{}\nlo{} accuracy.
This is the theoretical accuracy of the calculations of \citeres{Grazzini:2005vw,Frederix:2008vb}.
Including also the functions $g^{(3)}_{(N_1,N_2)}$ and ${\cal H}^{VV^\prime,(2)}$, together 
with the finite component at \nnlo{} (i.e.\ ${\cal O}(\as^2)$)
leads to full \nnll{}\plus{}\nnlo{} accuracy.
Using the recently computed two-loop amplitudes for $q{\bar q}\to VV^\prime$,
and the process independent relation of \citere{Catani:2013tia},
we are now able to present the complete result for the
transverse-momentum distribution of the vector-boson pair up to \nnll{}\plus{}\nnlo{} accuracy.
We point out
that the \nnll{}\plus{}\nnlo{} (\nll{}\plus{}\nlo{}) result includes the {\it full} \nnlo{} (\nlo{})
perturbative contribution in the entire $\pt$ range.
In particular,  the \nnlo{} (\nlo{}) result for the double differential cross section
$d{\sigma}/{(dM^2 dy)}$ is exactly recovered upon integration
over \pt{} of the differential cross section at \nnll{}\plus{}\nnlo{}
(\nll{}\plus{}\nlo{}) accuracy.

We conclude this Section by adding a few comments on the way in which our calculation
is actually performed.
The practical implementation of \eqn{resplusfin} is done in
the numerical program {\sc Matrix}\footnote{{\sc Matrix} is the abbreviation of 
``{\sc Munich} Automates qT subtraction and Resummation to Integrate X-sections", 
by M. Grazzini, S. Kallweit, D. Rathlev, M. Wiesemann. In preparation.}, 
which is an extension of the numerical program applied in the
\nnlo{} calculations of \citeres{Grazzini:2013bna,Cascioli:2014yka,Gehrmann:2014fva,Grazzini:2015nwa} 
and based on a combination of the $q_T$-subtraction formalism \cite{Catani:2007vq} 
with the \munich{}\footnote{\munich{} is the abbreviation of 
``MUlti-chaNnel Integrator at Swiss~(CH) precision''---an automated parton level \nlo{} 
generator by S.~Kallweit. In preparation.} code.
Since already performed within the $q_T$-subtraction formalism,
the extension of these calculations to compute the resummed cross section is conceptually quite straightforward, and is obtained by replacing the hard-collinear terms in the fixed-order 
computation by the proper all-order resummation formula of \eqn{resum}.
This procedure is the same that was
applied to perform the \nll{}\plus{}\nlo{} calculations for \ww{} \cite{Grazzini:2005vw}
and \zz{} \cite{Frederix:2008vb} production, and the \nnll{}\plus{}\nnlo{} calculation
for Higgs boson production of \citere{deFlorian:2012mx}.

To obtain the numerical results presented here,
the resummed component of \eqn{resum} is evaluated with an extension
of the numerical program used for the calculation of Higgs production \cite{deFlorian:2012mx}, based on the earlier computations of \citeres{Bozzi:2005wk,Bozzi:2007pn}.
The hard-collinear coefficients are obtained by exploiting the implementation of
the corresponding virtual amplitudes for the production of on-shell \zz{} and \ww{} pairs in \citere{Cascioli:2014yka} and \citere{Gehrmann:2014fva}, respectively, and the knowledge of the collinear coefficients relevant to quark-initiated processes \cite{Catani:2012qa}.\footnote{We note that an independent computation of these coefficients in the framework of Soft Collinear Effective Theory has been presented in \citeres{Gehrmann:2012ze,Gehrmann:2014yya}.}

The finite component of \eqn{resfin} is obtained from an \nlo{} calculation of $VV^\prime$+jet, computed with the \munich{} code, which
provides a fully automated implementation of the Catani--Seymour dipole 
formalism \cite{Catani:1996jh,Catani:1996vz}
as well as an interface to the one-loop generator \openloops{} 
\cite{Cascioli:2011va} to obtain all required (spin and color-correlated) 
tree-level and one-loop amplitudes.
For the numerically stable evaluation of tensor integrals we rely on the \Collier{} library~\cite{Denner:2014gla}, which is based on the Denner--Dittmaier reduction techniques~\cite{Denner:2002ii,Denner:2005nn} and the scalar integrals of~\cite{Denner:2010tr}.
To deal with problematic phase-space points, \openloops{} provides a rescue system using the quadruple-precision implementation of the OPP method in \CutTools{} \cite{Ossola:2007ax}, involving scalar integrals from \OneLOop{} \cite{vanHameren:2010cp}.

\section{Results}
\label{sec:results}

In this Section we present our results for the resummed 
transverse-momentum distributions of \ww{} and \zz{} pairs.
We compare our \nnll{}\plus{}\nnlo{} 
predictions to the results at the \nll{}\plus{}\nlo{}, and 
discuss the corresponding theoretical uncertainties.
Additionally, we 
also study the rapidity dependence of the \pt{} cross section as 
well as the \pt{}-veto efficiency.

For the \ew{} couplings we use the so-called $G_\mu$ scheme,
where the input parameters are $G_F$, $m_W$, $m_Z$. In particular we 
set 
$G_F = 1.16639\times 10^{-5}$~GeV$^{-2}$, $m_W=80.399$ GeV,
$m_Z = 91.1876$~GeV. 
We use the NNPDF3.0 sets of parton distribution functions (PDFs) 
\cite{Ball:2014uwa} with $\as(m_Z)=0.118$.
At \nll{}\plus{}\nlo{} and \nnll{}\plus{}\nnlo{} the running of $\as$ is evaluated
at two- and three-loop order, respectively.
For \zz{} production we consider $N_f=5$ massless quarks/antiquarks.
For \ww{} production we make use of the \fs{4},
which allows us to split off all contributions related to bottom-quark 
final states in order to remove the $t\bar t$ 
and $Wt$ contamination from our computation.
This is straightforward in the \fs{4}, because the $\ww b\bar{b}$ process is 
separately finite.

We consider proton-proton collisions at $\sqrt{s}=8$ TeV.
The central values of the 
factorization and renormalization scales are set 
to $\muF=\muR=\mu_0=2\,m_V$. The choice of the central resummation 
scale $Q_0$ is discussed in the next subsection.

\subsection{Choice of the central resummation scale}
\label{sec:resscale}

\begin{figure}[t]
\begin{center}
\includegraphics[trim = 10mm -5mm 0mm 0mm, height=.36\textheight]{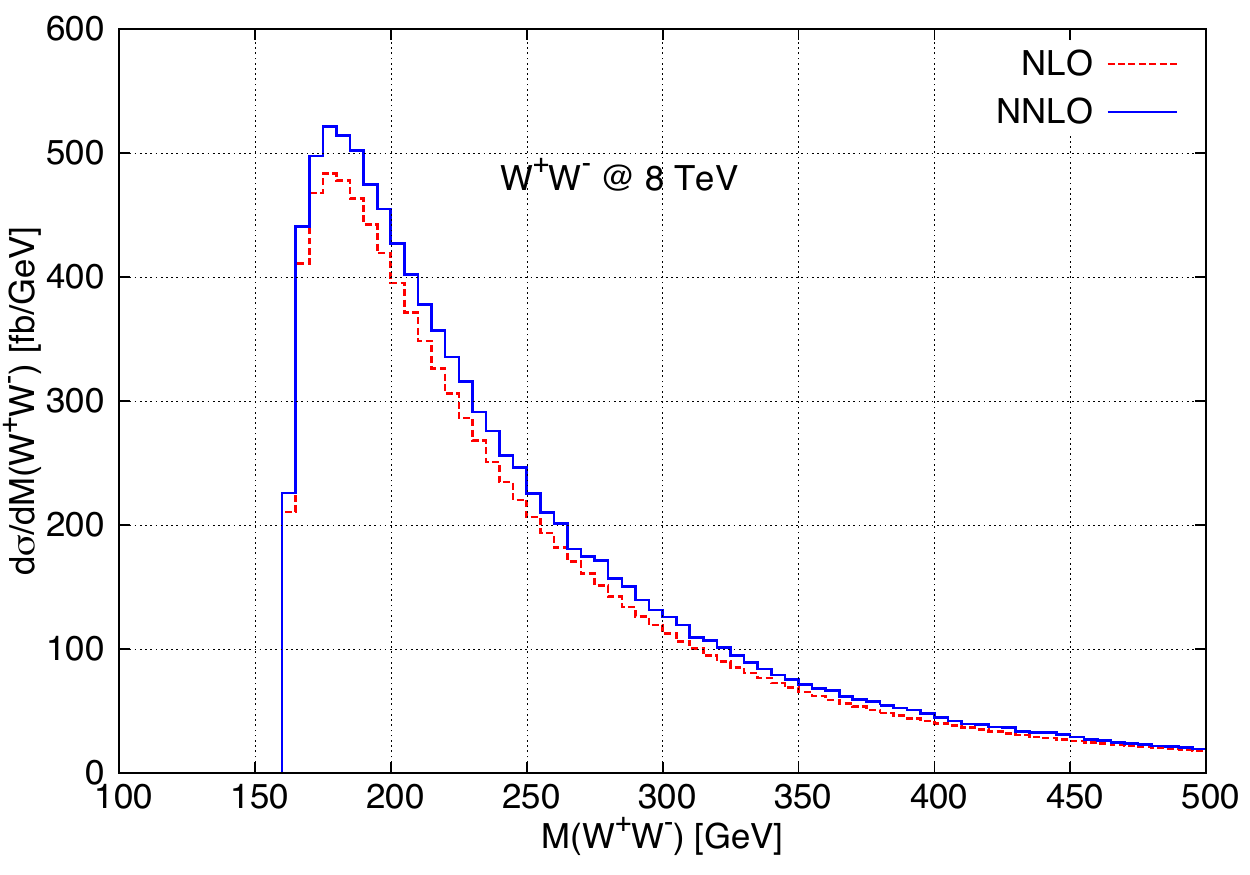}
    \parbox{.9\textwidth}{%
      \caption[]{\label{fig:mWW}{\sloppy Invariant mass ($M_{WW}$) distribution in \ww{} pair production at \nlo{} (red, dashed) and \nnlo{} (blue, solid).}  }}
\end{center}
\end{figure}

As discussed in \refse{sec:resum}, the resummation scale $Q$ is the scale entering the large logarithmic
terms we are resumming (see Eq.~(\ref{logpar})), and it plays the role of the scale up to which resummation is effective.
In on-shell Higgs \cite{Bozzi:2005wk} and vector-boson \cite{Bozzi:2010xn} production, the scale is typically chosen equal to half the mass
of the heavy boson (i.e.\ $Q=m_H/2$ for Higgs production and $Q=m_V/2$ in the case of single vector-boson production).
Higher values of the scale lead to a worse matching at high \pt{}. The natural extension of this choice for vector-boson
pair production is a dynamical resummation scale $Q=M_{VV}/2$, since $M_{VV}$ is the hardness of the process. This is indeed the choice that
was adopted in the calculations of Refs.~\cite{Grazzini:2005vw,Frederix:2008vb}.

The following considerations apply both to \zz{} and \ww{} production, so we will focus on \ww{} production from now on.
In \fig{fig:mWW} we consider the invariant mass distribution of the \ww{} pair at \nlo{} and \nnlo{}. We see that the distribution
is strongly peaked in the threshold region, and that it quickly decreases as $M_{WW}$ increases. As a consequence, for most of the \ww{} events, $M_{WW}\gtrsim 2m_W$.

\begin{figure}
\begin{center}
\includegraphics[trim = 10mm 0mm 0mm 0mm, height=.35\textheight]{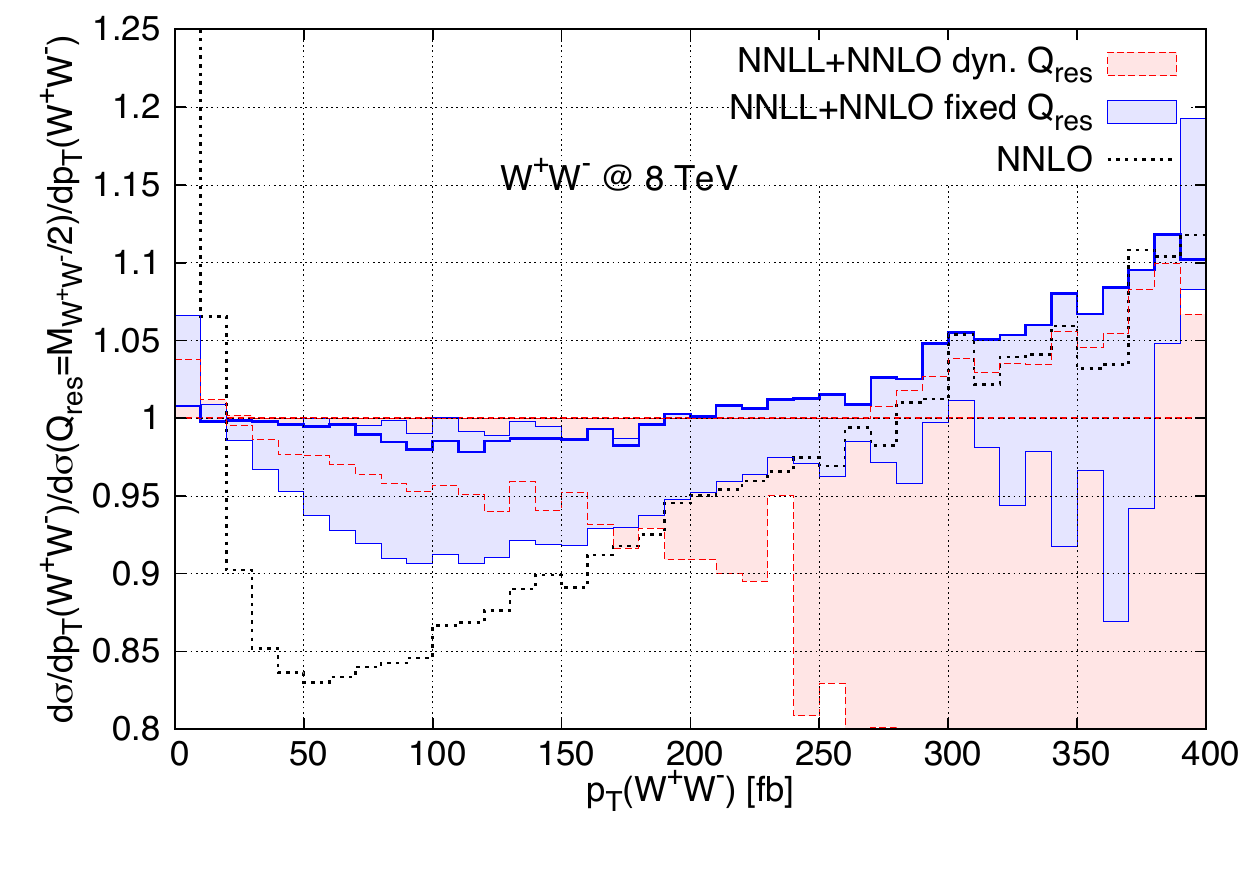}
    \parbox{.9\textwidth}{%
      \caption[]{\label{fig:fixvsdyn}{\sloppy \nnll{}\plus{}\nnlo{} transverse-momentum distribution of the \ww{} pair with a fixed scale $\Qres{}=m_W$ normalized to the same cross section with a dynamical scale $\Qres{}=M_{WW}/2$. The bands are obtained by variation of the resummation scales in the numerator by a factor of two around the central scale.  For reference, we show the fixed-order \nnlo{} curve with the same normalization.}  }}
\end{center}
\end{figure}

We can compare the transverse-momentum distributions obtained with a dynamical resummation scale $Q=M_{WW}/2$, and a fixed resummation scale $Q=m_W$.
In \fig{fig:fixvsdyn} we show the 
ratio (blue, solid curve) of the $Q=m_W$ result over the $Q=M_{WW}/2$ result. The 
bands are obtained by varying the resummation scale around the central value by a 
factor of two. Considering the ratio of the central curves for $\pt{}\lesssim 250$\,GeV, 
the differences between a fixed and a dynamical scale are extremely small and remain at the 1-2\% level
over the whole range. In this region of transverse momenta
the uncertainty bands obtained with the two choices overlap and are similar in size.
In fact, since $Q=m_W$ leads to slightly larger uncertainties, it appears 
to be the more conservative choice.
Therefore, we can conclude that either choice of 
the resummation scale is perfectly valid and indeed consistent with each other as 
expected from the discussion of the invariant mass distribution.

Looking further at the comparison of the high-\pt{} tails in \fig{fig:fixvsdyn} ($\pt{}
\gtrsim 250$\,GeV), we observe a very well known feature \cite{Bozzi:2005wk,Bozzi:2010xn,deFlorian:2012mx,Harlander:2014hya,Harlander:2014uea} of the applied 
matching procedure, namely the fact that for large values of the resummation 
scale the fixed-order cross section (black dotted curve) is not recovered in the tail of 
the distribution. It is important to recall that transverse-momentum resummation
is supposed to improve the perturbative expansion in the low-\pt{} region.
At large \pt{}, any large dependence on the 
resummation scale is necessarily artificial and an unwanted remnant 
of the matching procedure. This behaviour is precisely what 
we observe for the dynamical scale choice $Q=M_{WW}/2$ in \fig{fig:fixvsdyn}.
With this choice, in fact, the resummed result loses predictivity, as its uncertainty becomes
increasingly large. 
By contrast, a fixed resummation scale $Q=m_W$, which is always smaller 
than $Q=M_{WW}/2$, eventually leads to a more consistent high-\pt{}
behaviour of the resummed prediction.

Based on the above results, we make $Q_0=m_V$ our default choice of the 
resummation scale in what follows.

\subsection{Inclusive transverse-momentum distribution}
\label{sec:inclpt}

\begin{figure}
\begin{center}
\includegraphics[trim = 10mm -5mm 0mm 0mm, height=.37\textheight]{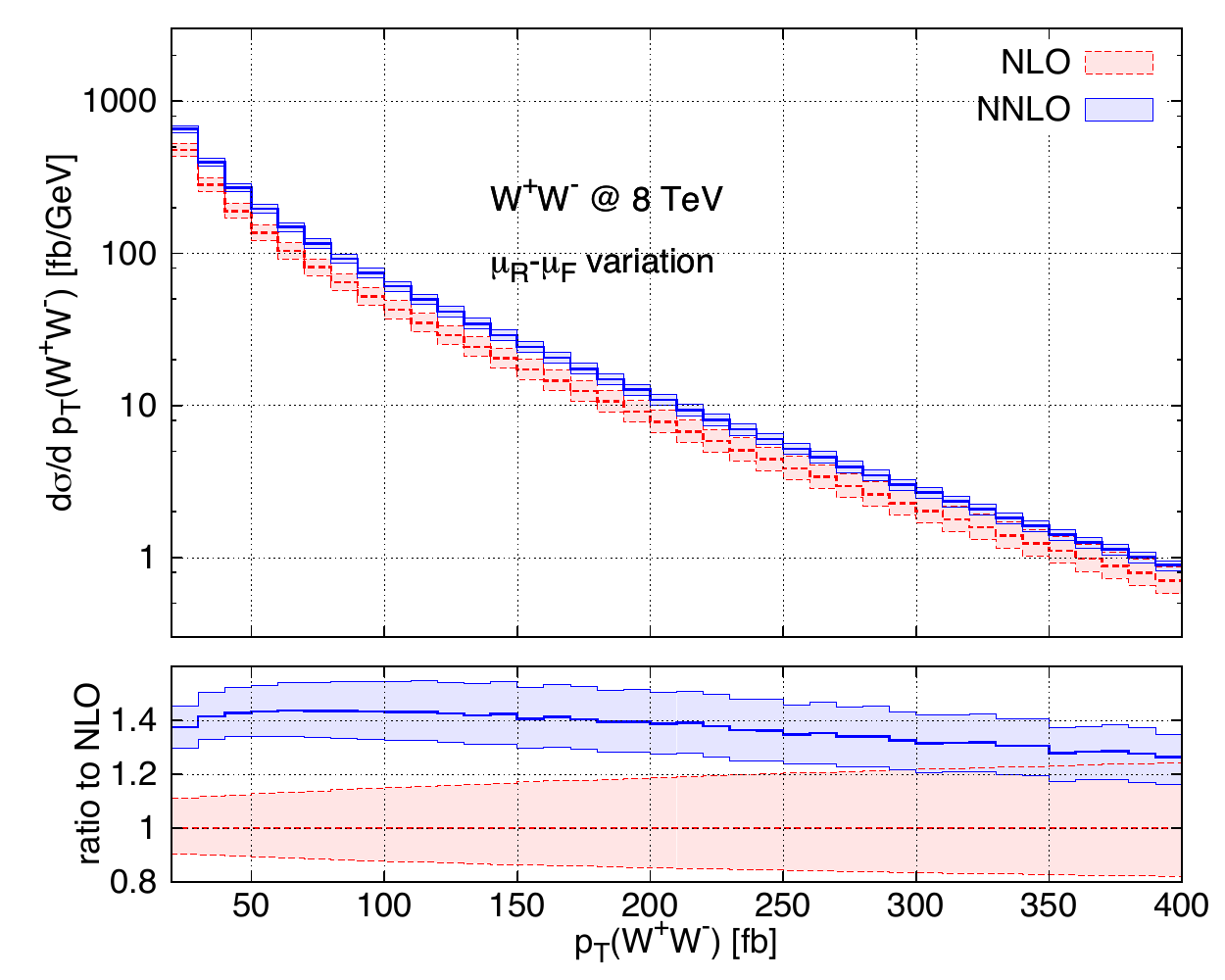}
\parbox{.9\textwidth}{%
      \caption[]{\label{fig:fixedorder}{
            \sloppy  Transverse-momentum distribution of the \ww{} pair
          at \nlo{} (red, dashed) and \nnlo{} (blue, solid); thick lines: central scale choices; bands: scale uncertainty 
          from $\muF{}$ and $\muR{}$ obtained as described in the text. Lower inset: results normalized to the \nlo{} prediction at central values of the scales.
}}}
\end{center}
\end{figure}

We now present our resummed predictions for the inclusive transverse-momentum spectrum
of the vector-boson pair and compare them with the corresponding fixed-order results.
We concentrate on \ww{} production since we observe no saliently different 
features in the \zz{} case. For completeness, we provide 
the corresponding reference prediction with uncertainties for \zz{} below.

Before presenting our resummed predictions,
we recall the well known fixed-order results at ${\cal O}(\as)$ and ${\cal O}(\as^2)$ \cite{Dittmaier:2007th,Campbell:2007ev,Dittmaier:2009un}.
In \fig{fig:fixedorder} we show the \nlo{} and \nnlo{} distributions together with their perturbative uncertainties. The uncertainty bands are obtained by varying $\muF$ and $\muR$ in the range $m_W\leq \{\muF,\muR\} \leq 4 m_W$
with the constraint $0.5\leq \muF/\muR\leq 2$. The lower inset shows the same results normalized to the central
\nlo{} curve.
The \nnlo{} effects range from about $40\%$ at $\pt\sim 50$ GeV to about $30\%$ at $\pt\sim400$ GeV. The \nlo{} (\nnlo{}) uncertainty ranges from about $\pm 15\%$ ($\pm 10\%$) at $\pt\sim 50$ GeV to about $\pm 20\%$ ($\pm 8\%$) at $\pt\sim400$ GeV.
We note that the \nlo{} and \nnlo{} bands do not overlap in the region where $\pt\lesssim 300$ GeV.
This implies that, in this region of transverse momenta, the size of the band obtained through
scale variations at \nlo{} definitely underestimates the theoretical uncertainty.

\begin{figure}[p]
\begin{center}
\begin{tabular}{cc}
\hspace*{-0.17cm}
\includegraphics[trim = 7mm -5mm 0mm 0mm, height=.322\textheight,]{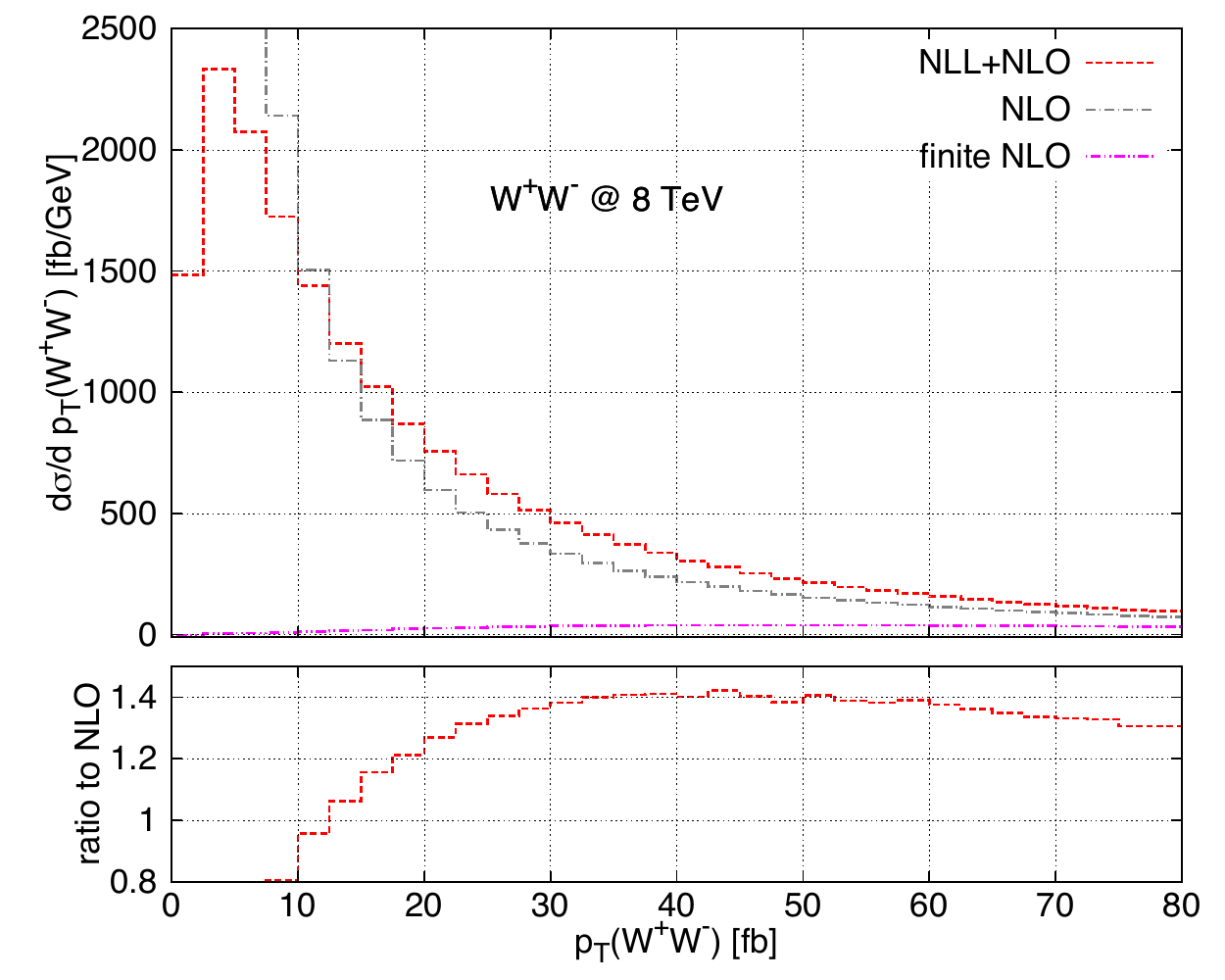} &
\includegraphics[trim = 7mm -5mm 0mm 0mm, height=.322\textheight]{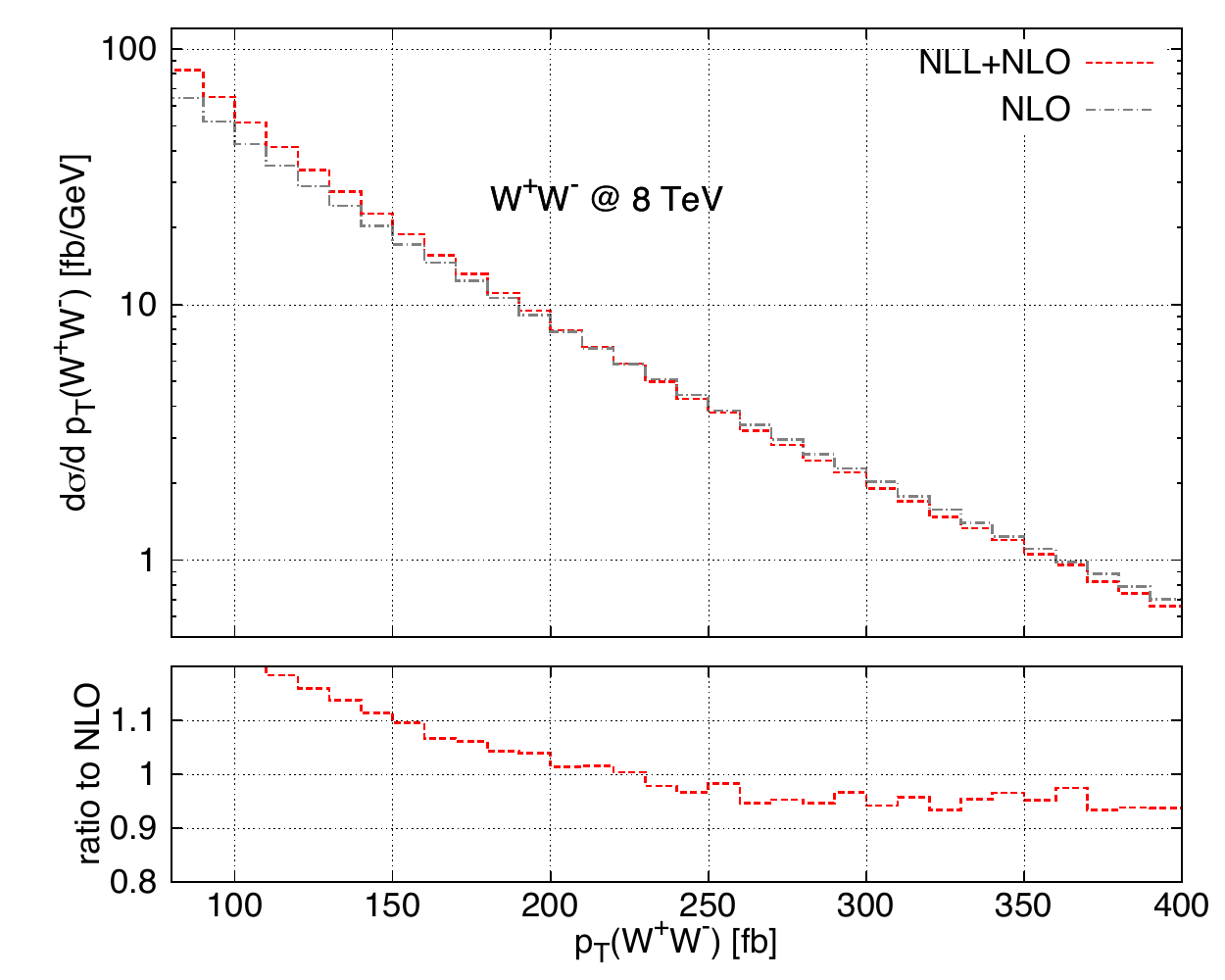} \\[-1em]
\hspace{0.6em} (a) & \hspace{1em}(b)
\end{tabular}\vspace{0.2cm}
  \parbox{.9\textwidth}{%
      \caption[]{\label{fig:matchingNLO}{\sloppy The transverse-momentum spectrum of the \ww{} pair at \nll{}\plus{}\nlo{} (a) in the low-\pt{} region and (b) at high transverse momenta. The \nll{}\plus{}\nlo{} result (red, dashed) is compared to the fixed-order \nlo{} prediction (grey, dash-dotted) and to the finite component of \eqn{resplusfin} (magenta, dash-double dotted). The lower insets show the \nll{}\plus{}\nlo{} result normalized to \nlo{}.}}}
\end{center}
\begin{center}
\begin{tabular}{cc}
\hspace*{-0.17cm}
\includegraphics[trim = 7mm -5mm 0mm 0mm, height=.322\textheight,]{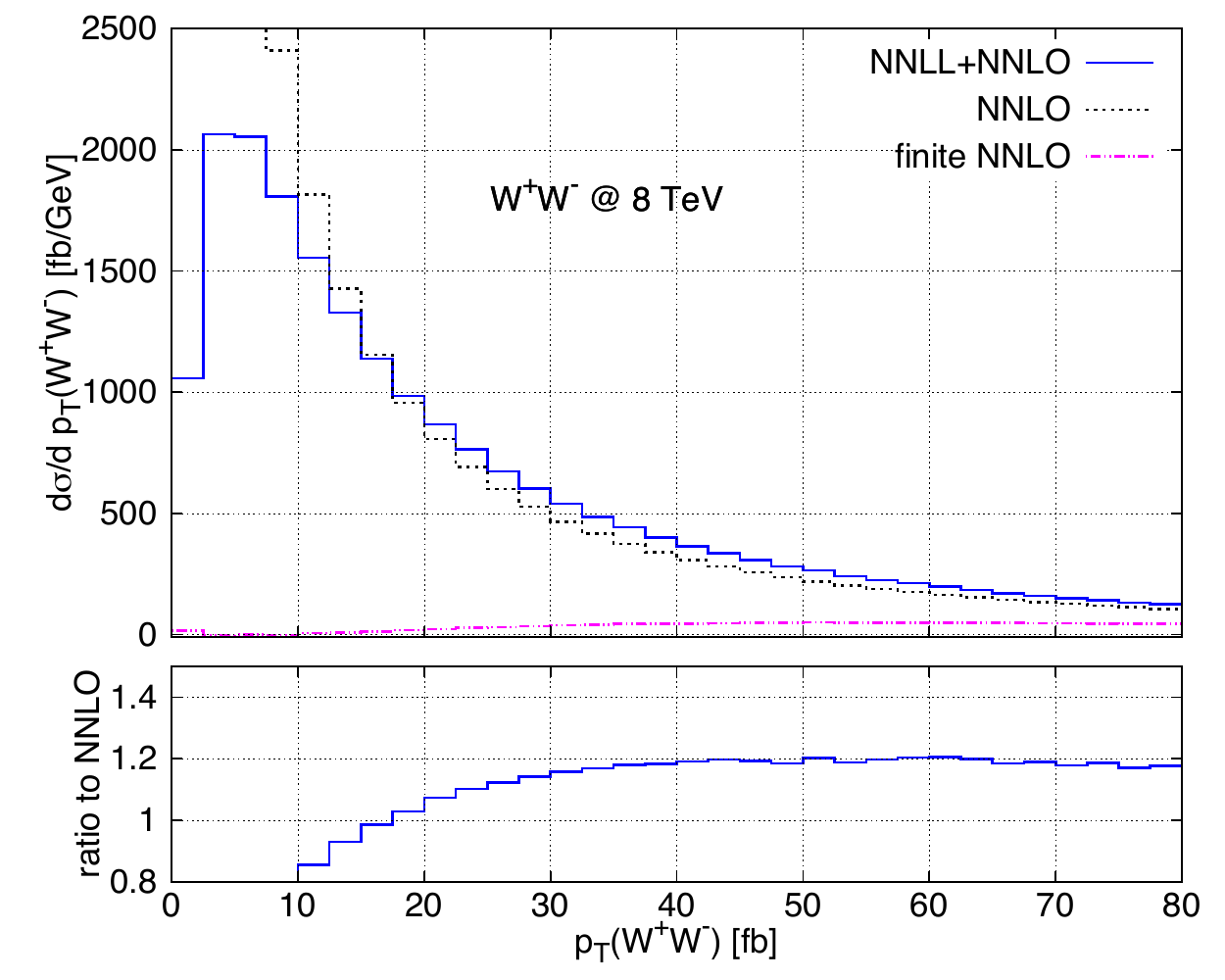} &
\includegraphics[trim = 7mm -5mm 0mm 0mm, height=.322\textheight]{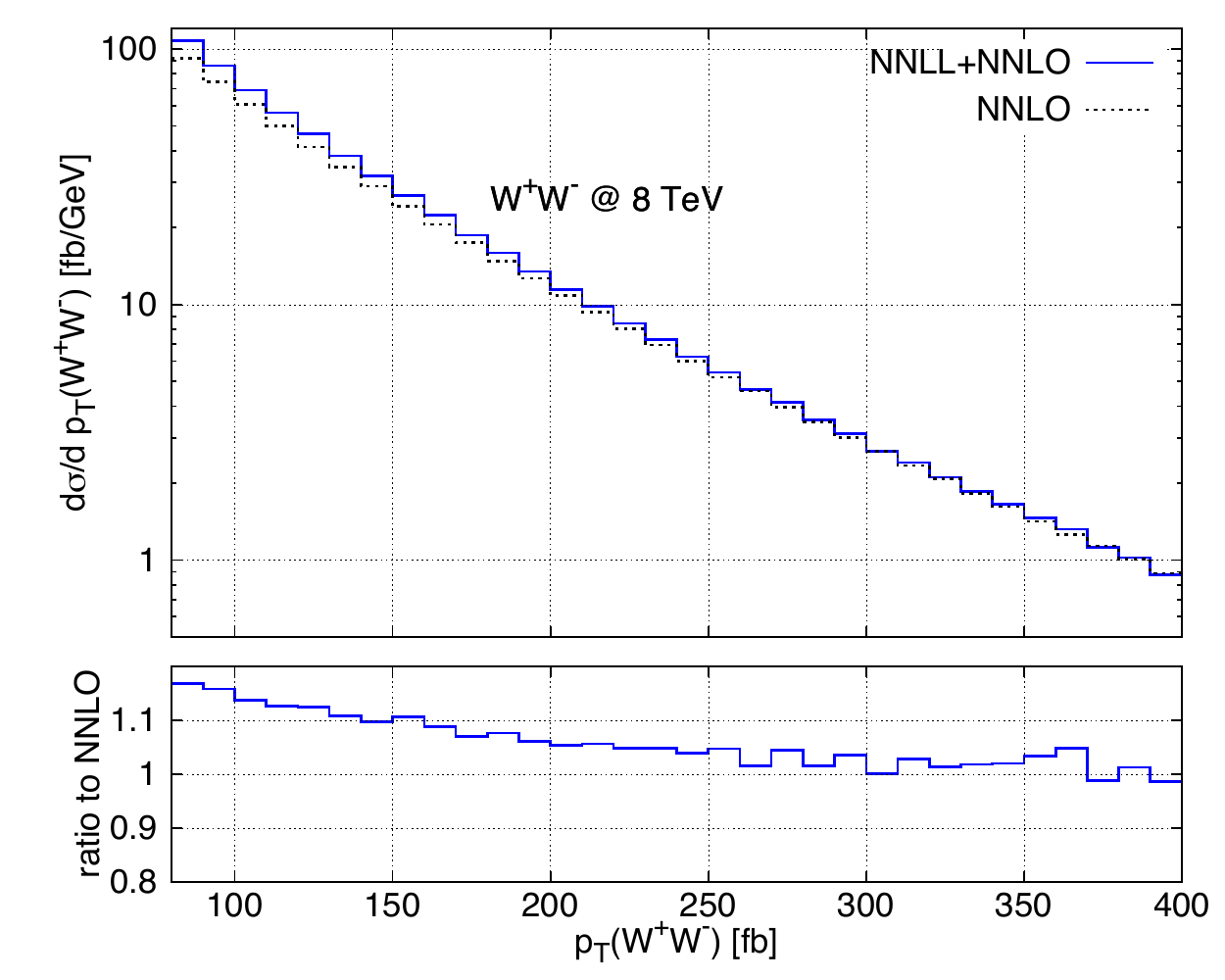} \\[-1em]
\hspace{0.6em} (a) & \hspace{1em}(b)
\end{tabular}\vspace{0.2cm}
  \parbox{.9\textwidth}{%
      \caption[]{\label{fig:matchingNNLO}{\sloppy The transverse-momentum spectrum of the \ww{} pair at \nnll{}\plus{}\nnlo{} (a) in the low-\pt{} region and (b) at high transverse momenta. The \nnll{}\plus{}\nnlo{} result (red, dashed) is compared to the fixed-order \nnlo{} prediction (grey, dash-dotted) and to the finite component of \eqn{resplusfin} (magenta, dash-double dotted). The lower insets show the \nnll{}\plus{}\nnlo{} result normalized to \nnlo{}.}}}
\end{center}
\end{figure}

We now move on to the resummed results. In \fig{fig:matchingNLO}\,(a) the \nll{}\plus{}\nlo{} spectrum is compared to the fixed-order \nlo{} result and to the
finite component of the resummed cross section (see \eqn{resplusfin}) in the region between $0$ and $80$\,GeV. As expected, the \nlo{} diverges to $+\infty$ as $p_T\to 0$,
while the resummation provides a physically well behaved spectrum down to low values of \pt{}, which exhibits a kinematical peak at $p_T\sim 4$ GeV. The finite component contributes less than $1\%$ in the peak region, where the result is dominated by resummation, and it increases to $\sim18\%$ at $p_T=50$ GeV. The lower 
inset shows the \nll{}\plus{}\nlo{} result normalized to \nlo{}. In \fig{fig:matchingNLO}\,(b) the region between $80$ and $400$ GeV is displayed. We see that even at large values of \pt{} the \nll{}\plus{}\nlo{} resummed result does not match very well the fixed-order \nlo{} result, with a difference of about $5$\%. 

The analogous results at \nnll{}\plus{}\nnlo{} are shown in \fig{fig:matchingNNLO}. The \nnlo{} has an unphysical (divergent) behaviour as $p_T\to 0$, whereas the resummed spectrum is well behaved, with a slightly harder peak with respect to the \nll{}\plus{}\nlo{}.
The finite component contributes less than $1\%$ in the peak region,
increasing to $\sim19\%$ at $p_T=50$ GeV.
Comparing the right panels of \fig{fig:matchingNLO} and \fig{fig:matchingNNLO}, we see that the quality of the matching at high \pt{} is significantly improved when going from \nll{}\plus{}\nlo{} to \nnll{}\plus{}\nnlo{}, and we find that this behaviour is indeed preserved up to very high transverse momenta.
The \nnll{}\plus{}\nnlo{} result thus gives a prediction with uniform accuracy from small to very large transverse momenta and, in fact, provides a sufficiently large 
region where a hard switching to the fixed-order result is feasible.
We point out that, thanks to our unitarity constraint, both at \nll{}\plus{}\nlo{} and at \nnll{}\plus{}\nnlo{} the integral of the resummed spectrum is in excellent agreement with the respective total cross sections; the differences are at the few-permille level.

\begin{figure}
\begin{center}
\hspace*{-0.15cm}
\begin{tabular}{cc}
\includegraphics[trim = 7mm -5mm 0mm 0mm, height=.324\textheight,]{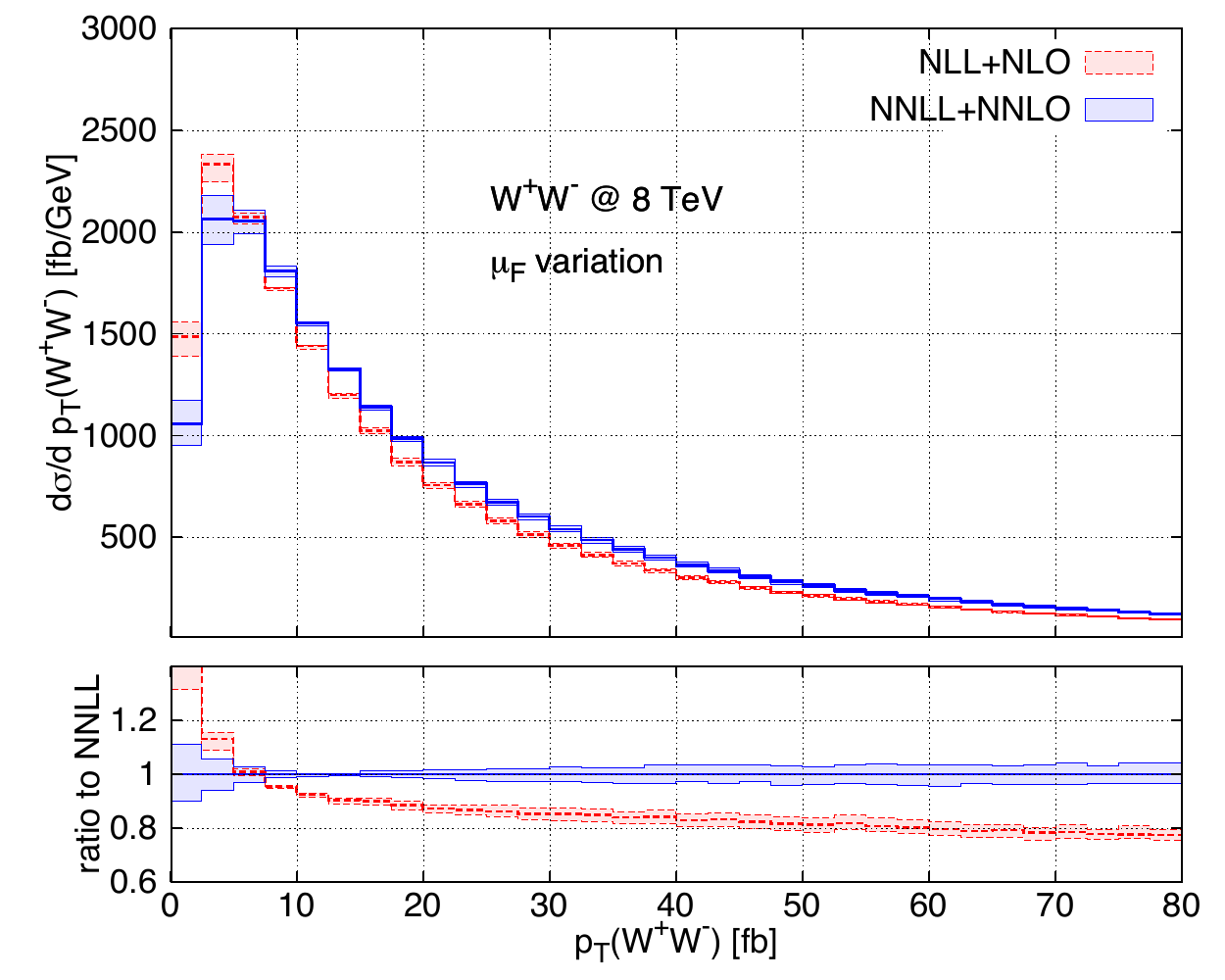} &
\includegraphics[trim = 7mm -5mm 0mm 0mm, height=.324\textheight]{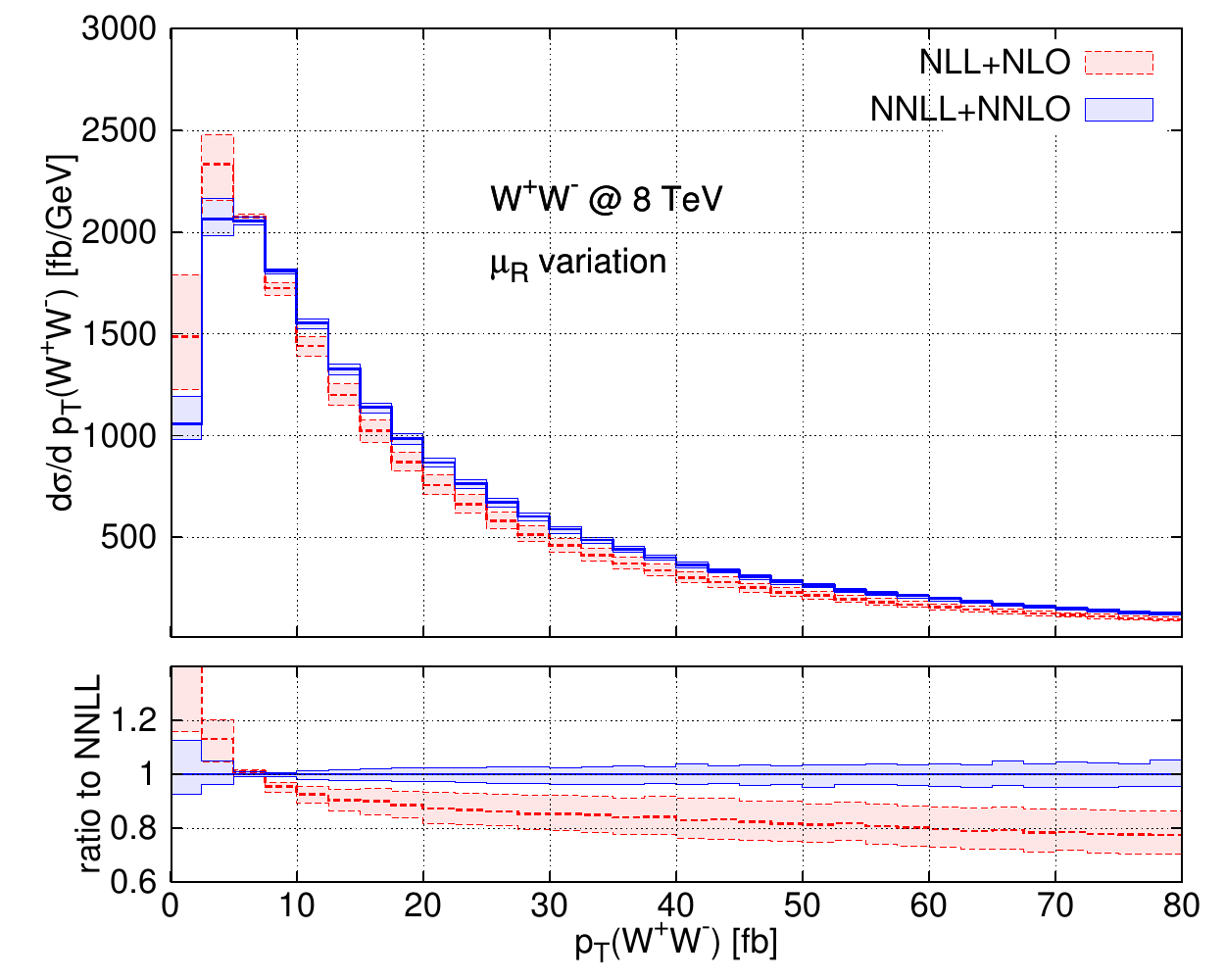} \\[-1em]
\hspace{0.6em} (a) & \hspace{1em}(b)
\end{tabular}\vspace{0.2cm}
  \parbox{.9\textwidth}{%
      \caption[]{\label{fig:mufmur}{\sloppy  \ww{} transverse-momentum distribution
          at the \nll{}\plus\nlo{} (red, dashed) and \nnll{}\plus\nnlo{}
          (blue, solid); thick lines: central scale choices; bands: uncertainty
          due to (a) $\muF{}$ variation and (b) $\muR{}$ variation; thin lines: borders of bands.
}}}
\end{center}
\end{figure}
\begin{figure}
\begin{center}
\begin{tabular}{cc}
\hspace*{-0.15cm}
\includegraphics[trim = 7mm -5mm 0mm 0mm, height=.324\textheight,]{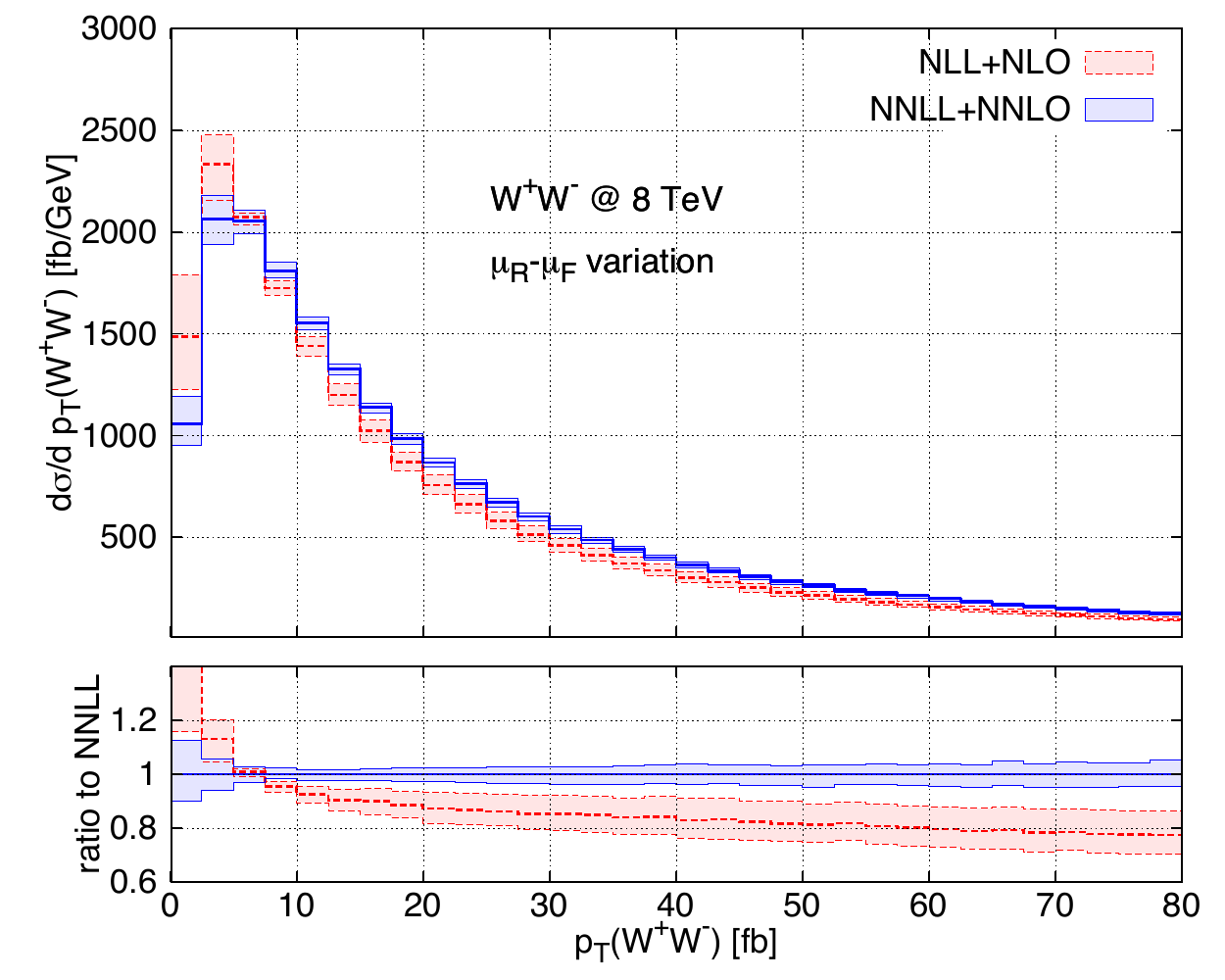} &
\includegraphics[trim = 7mm -5mm 0mm 0mm, height=.324\textheight]{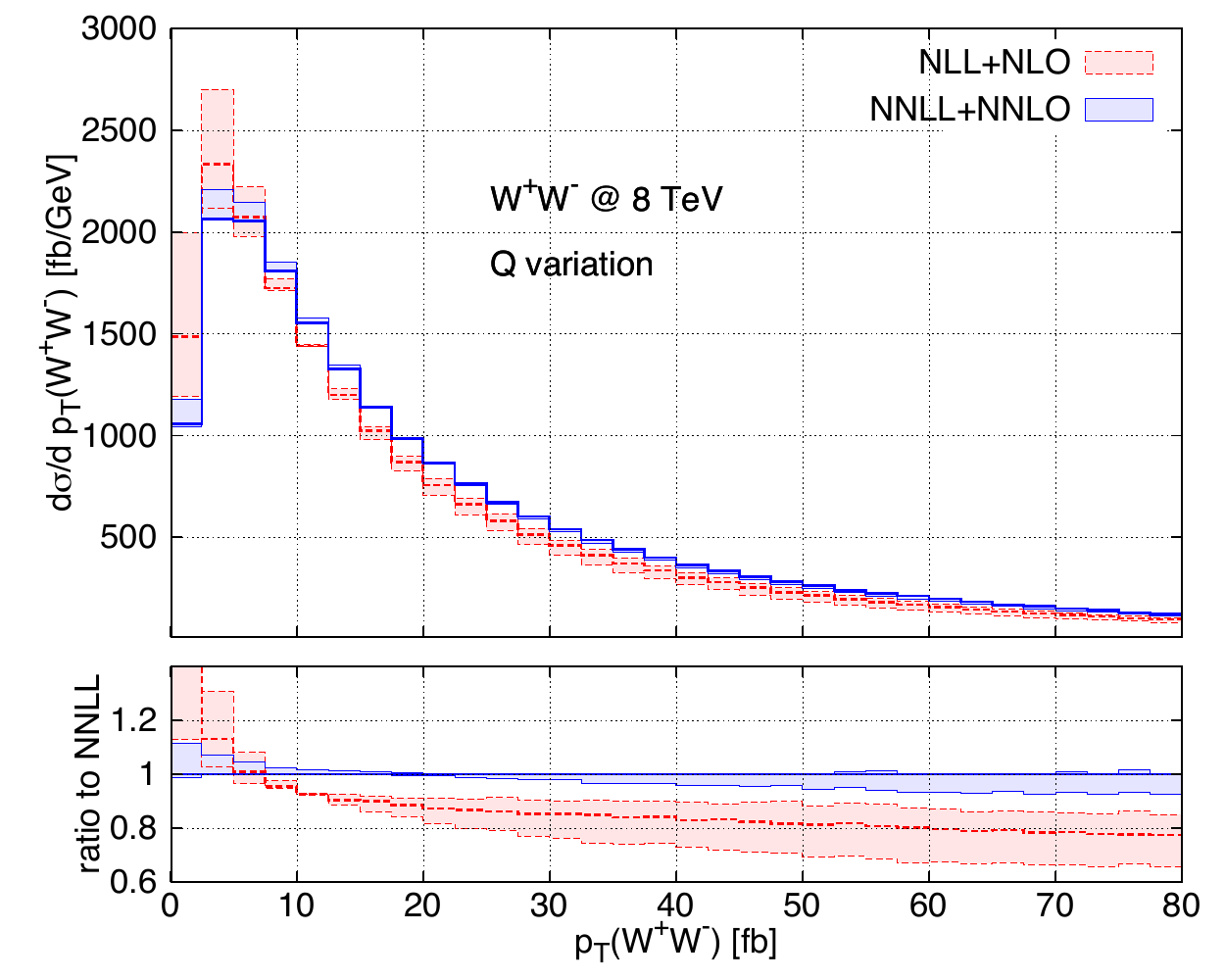} \\[-1em]
\hspace{0.6em} (a) & \hspace{1em}(b)
\end{tabular}\vspace{0.2cm}
  \parbox{.9\textwidth}{%
      \caption[]{\label{fig:unc}{\sloppy  
          \ww{} transverse-momentum distribution
          at the \nll{}\plus\nlo{} (red, dashed) and \nnll{}\plus\nnlo{}
          (blue, solid); thick lines: central scale choices; bands: uncertainty
          due to (left) $\muF{},\muR{}$ variation and (right) $Q$ variation; thin lines: borders of bands.
}}}
\end{center}
\end{figure}

We now turn to the scale uncertainties of our resummed results. We start our discussion by separately considering
factorization and renormalization scale variations. In \fig{fig:mufmur} we compare the \nll{}\plus{}\nlo{} (red, dashed) and \nnll{}\plus{}\nnlo{} (blue, solid)
predictions with their uncertainty bands from $\muF$ and $\muR$ variations (left and right panel, respectively).
In both cases, the bands are obtained by varying the factorization (renormalization) scale by a factor of two around its central value, while keeping the other scales at their default values.
First of all, we notice that when going from \nll{}\plus{}\nlo{} to \nnll{}\plus{}\nnlo{} the \pt{} spectrum becomes harder.
Comparing with the results of Ref.~\cite{Grazzini:2005vw}, where the \nnll{} resummation was implemented without ${\cal O}(\as^2)$ matching,
we see that the increased hardness of the \pt{} spectrum is a combined effect of both features, i.e.\ \nnll{} resummation and
\nnlo{} matching at high \pt{}.

We note that neither in the case of the factorization
scale, nor in the case of the renormalization scale, the \nll{}\plus{}\nlo{} and \nnll{}\plus{}\nnlo{} bands overlap. Actually, in the case of the factorization scale, there is no reduction in scale dependence when going from \nll{}\plus{}\nlo{} to \nnll{}\plus{}\nnlo{}, and the uncertainty
slightly increases with the perturbative order, even if it is always well below $10\%$, except at very low \pt{}. 
The renormalization scale dependence instead exhibits the expected reduction when going from \nll{}\plus{}\nlo{} to \nnll{}\plus{}\nnlo{}.

In \fig{fig:unc} we present our resummed predictions with uncertainty bands obtained from simultaneous variations
of $\muF$ and $\muR$ (left panel) and the variation of the resummation scale $Q$ (right panel).
In the left panel the uncertainty bands are obtained by varying $\muF$ and $\muR$ as in \fig{fig:fixedorder}.
In the right panel the
resummation scale is varied in the range $m_W/2\leq Q\leq 2m_W$.
As in \fig{fig:mufmur} we see that the uncertainty bands do not overlap.
The uncertainty from $\muF$ and $\muR$ variations is $\pm 10-15\%$ at \nll{}\plus{}\nlo{} and is reduced to $8-10\%$ at \nnll{}\plus{}\nnlo{}.
At \nll{}\plus{}\nlo{} the resummation scale uncertainty is generally about $\pm 15\%$ except in the region of $p_T\sim 10$ GeV, where it shrinks to smaller values. We find that at \nnll{}\plus{}\nnlo{} the resummation scale uncertainty is reduced roughly by a factor of two in the region of transverse momenta considered in the figure.

\begin{figure}[p]
\begin{center}
\hspace*{-0.15cm}
\begin{tabular}{cc}
\includegraphics[trim = 7mm -5mm 0mm 0mm, height=.324\textheight,]{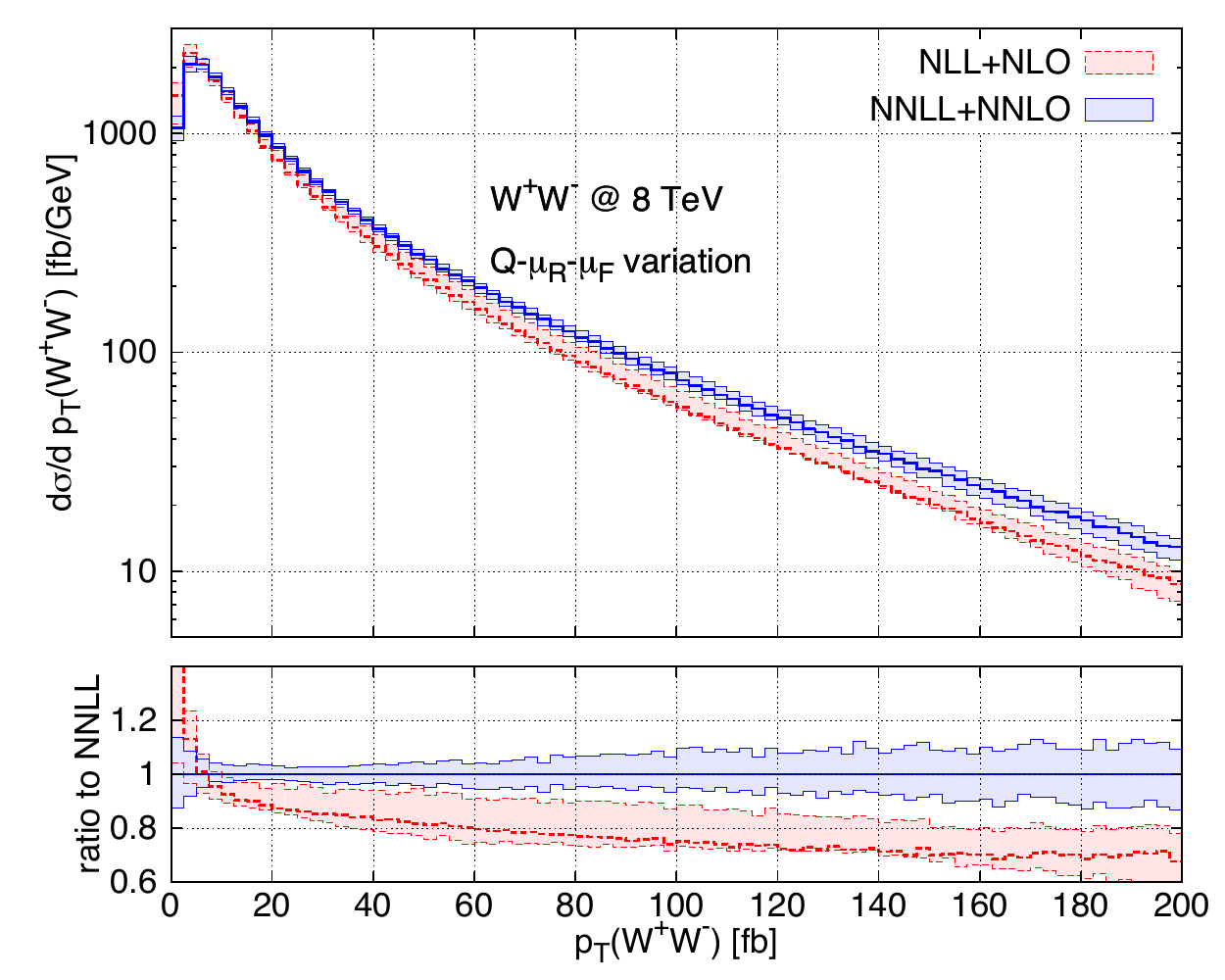} &
\includegraphics[trim = 7mm -5mm 0mm 0mm, height=.324\textheight]{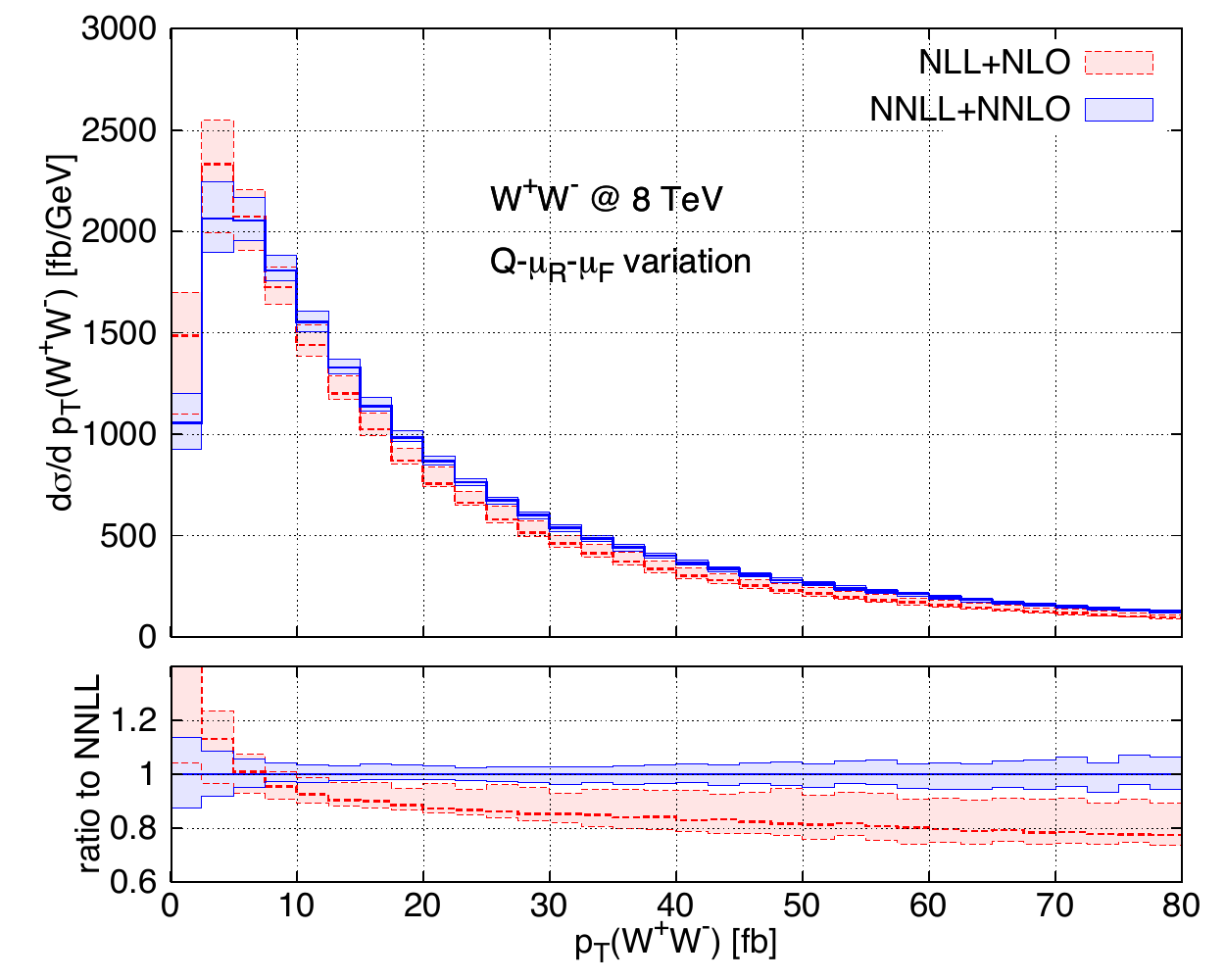} \\[-1em]
\hspace{0.6em} (a) & \hspace{1em}(b)
\end{tabular}\vspace{0.2cm}
  \parbox{.9\textwidth}{%
      \caption[]{\label{fig:bestpredictionww}{
            \sloppy  (a) Transverse-momentum distribution of the \ww{} pair
          at \nll{}\plus\nlo{} (red, dashed) and \nnll{}\plus\nnlo{}
          (blue, solid); thick lines: central scale choices; bands: uncertainty 
          from $\muF{}$, $\muR{}$ and $Q$ variations obtained as described in the text; thin lines: borders of bands. (b) detail of the low-\pt{} region.
}}}
\end{center}
\begin{center}
\hspace*{-0.15cm}
\begin{tabular}{cc}
\includegraphics[trim = 7mm -5mm 0mm 0mm, height=.324\textheight,]{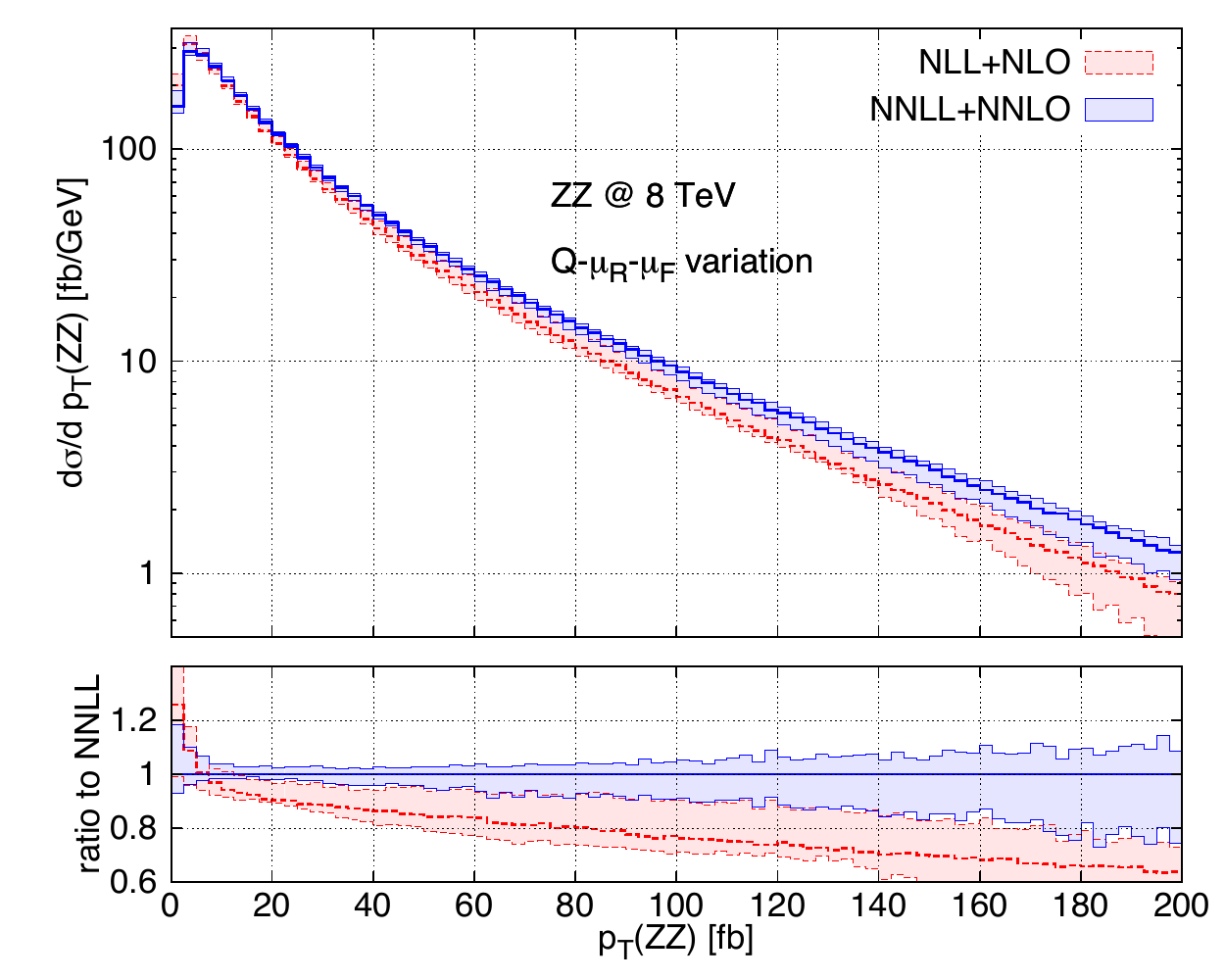} &
\includegraphics[trim = 7mm -5mm 0mm 0mm, height=.324\textheight]{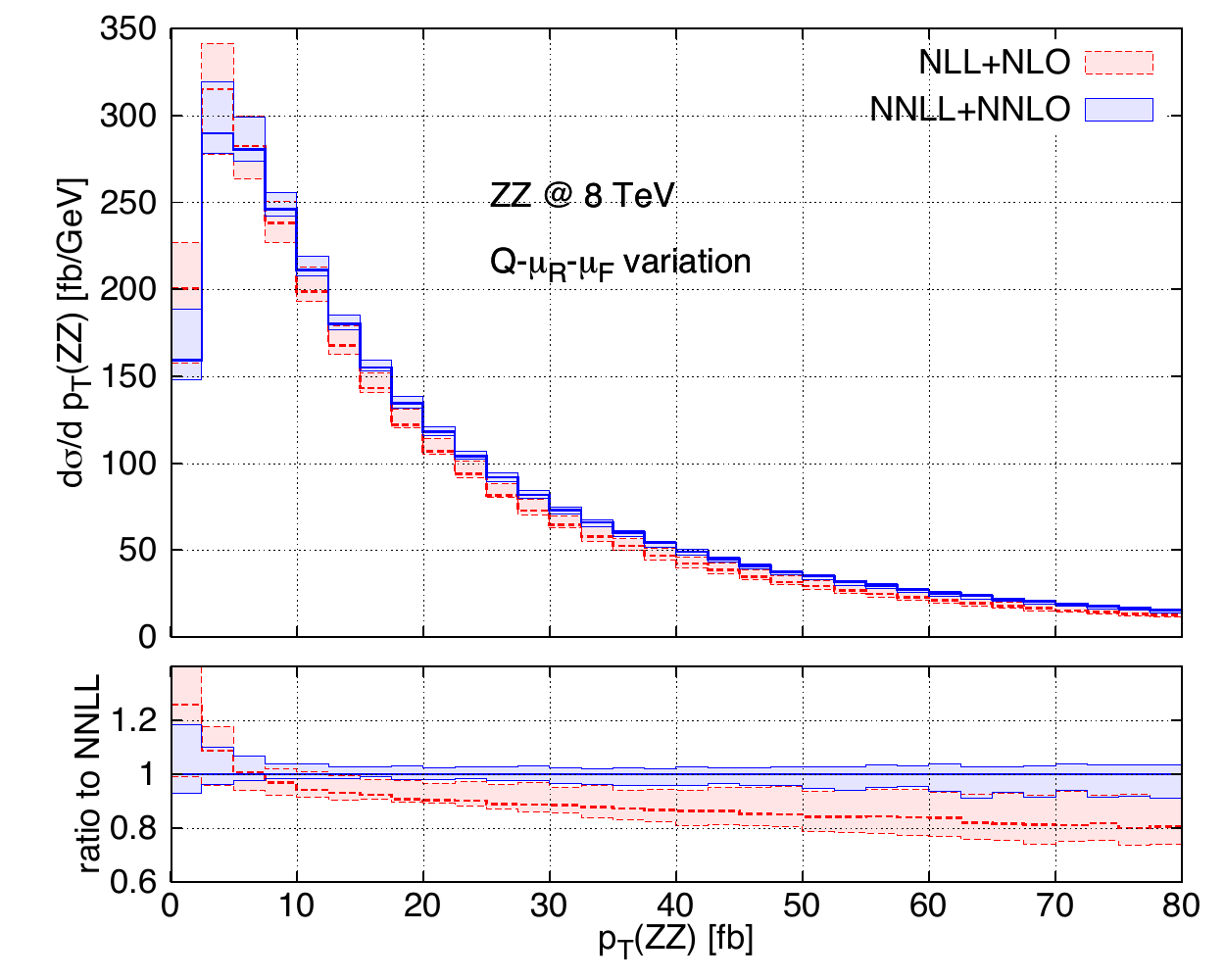} \\[-1em]
\hspace{0.6em} (a) & \hspace{1em}(b)
\end{tabular}\vspace{0.2cm}
  \parbox{.9\textwidth}{%
      \caption[]{\label{fig:bestpredictionzz}{
            \sloppy  Same as \fig{fig:bestpredictionww}, but for \zz.
}}}
\end{center}
\end{figure}

In \figs{fig:bestpredictionww} and \ref{fig:bestpredictionzz}
we show our reference resummed prediction for \ww{} and \zz, respectively,
with an estimate of their full perturbative uncertainty.
In order to obtain a combined uncertainty from $\muF$, $\muR$ and $Q$ variations, we follow Ref.~\cite{Bozzi:2010xn} and
independently vary $\muF$, $\muR$ and $Q$ in the ranges $m_V\leq \{\muF,\muR\} \leq 4m_V$
and $m_V/2 \leq Q \leq 2m_V$ with the constraints $0.5\leq \muF/\muR\leq 2$ and $0.5\leq Q/\muR\leq 2$.
We recall that the constraint on $\muF/\muR$, which is the same as applied in \fig{fig:fixedorder} and \fig{fig:unc} (left), has the purpose
of avoiding large logarithmic contributions from the evolution of parton densities. Analogously, the constraint
on $Q/\muR$ avoids large logarithmic contributions in the expansion of the Sudakov form factor.

For \ww{} production the perturbative uncertainty at \nnll{}\plus{}\nnlo{} (\nll{}\plus{}\nlo{}) is about $\pm 8\%$ ($\pm 12\%$)
at the peak, it decreases to about $\pm 3\%$ ($\pm 5\%$) at $p_T=20$ GeV,
and it increases again to $\pm 10\%$ ($\pm 15\%$) at $p_T=200$ GeV. In the high-\pt{} region, the difference between the \nnll{}\plus{}\nnlo{} and \nll{}\plus{}\nlo{} predictions is driven by the \nnlo{} effects, which increase the \nlo{} result by about $30\%$.

For \zz{} production the uncertainties have essentially the same pattern in the small- and
intermediate-\pt{} region, while at high \pt{} they are larger than for \ww{} production,
reaching about $\pm 17\%$ at \nnll{}\plus{}\nnlo{} for $p_T=200$ GeV.
We have checked that this effect is entirely driven by the resummation-scale dependence.
As previously pointed out, this behaviour is not particularly worrying since,
in the large-\pt{} region, the resummed results should be replaced by the corresponding fixed-order prediction.
Also in the \zz{} case the 
large enhancement of the \nnll{}\plus\nnlo{} distribution in the high-\pt{} tail 
stems from the fixed-order cross section.

\subsection{Rapidity dependence of the transverse-momentum distribution}
\label{sec:rap}

\begin{figure}[h]
\begin{center}
\begin{tabular}{cc}
\includegraphics[trim = 7mm -5mm 0mm 0mm, height=.283\textheight,]{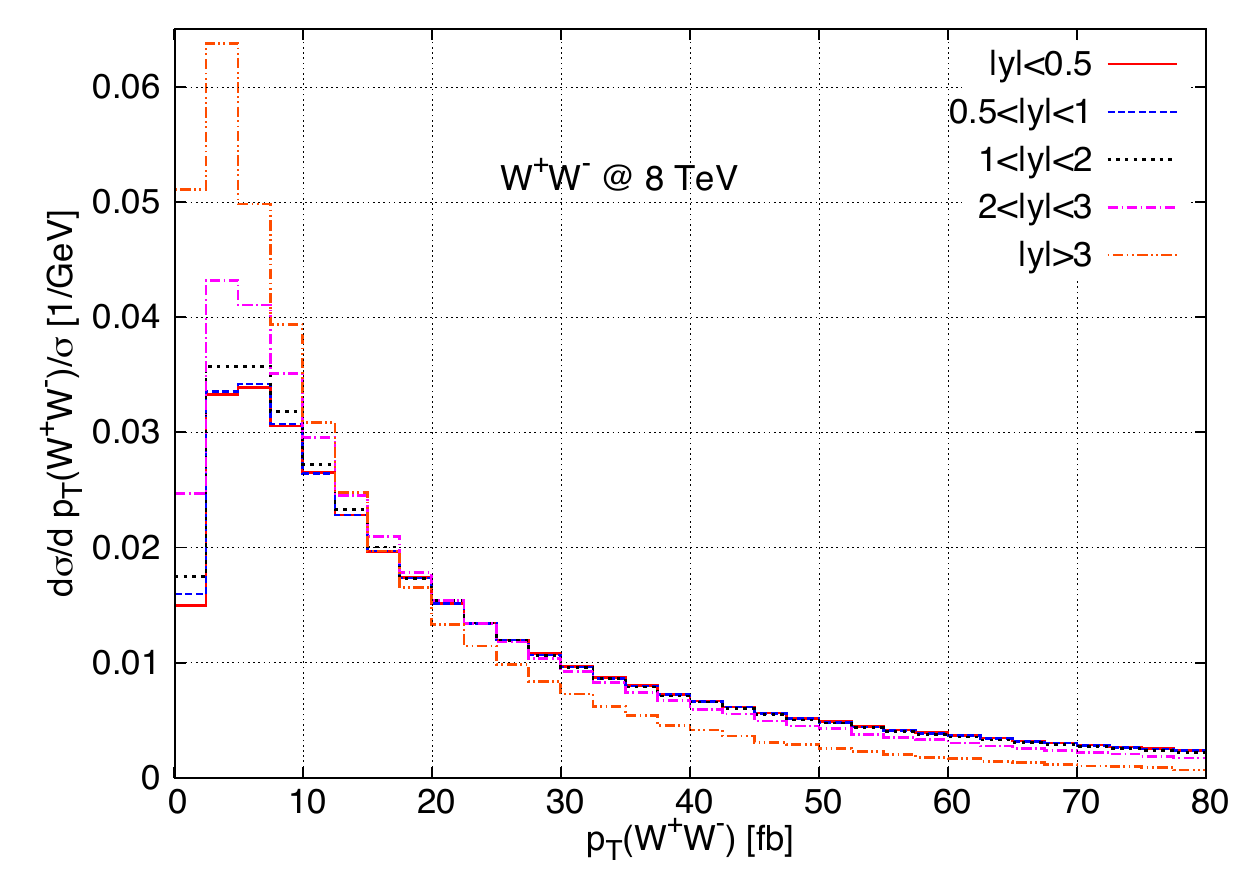} &
\includegraphics[trim = 7mm -5mm 0mm 0mm, height=.283\textheight]{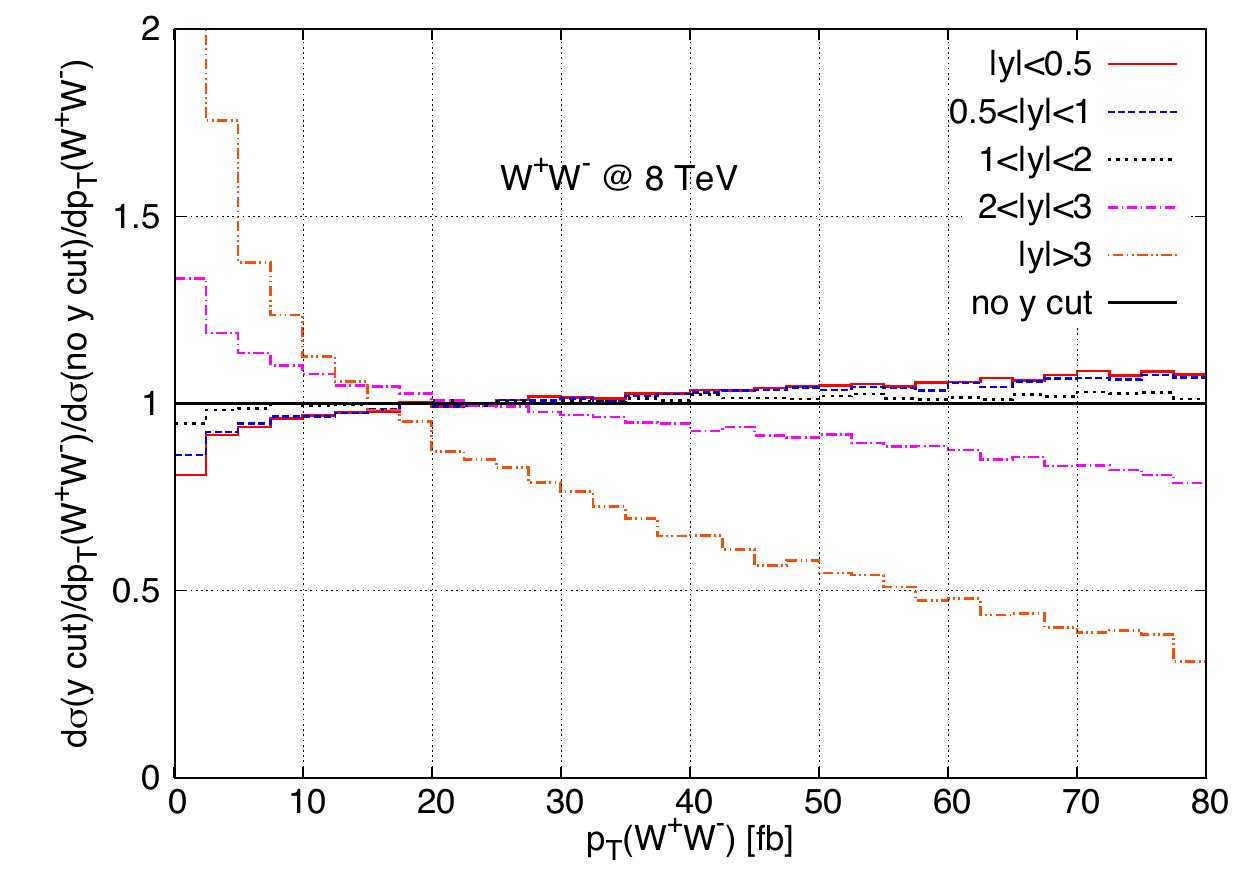} \\[-1em]
\hspace{0.6em} (a) & \hspace{1em}(b)
\end{tabular}\vspace{0.2cm}
  \parbox{.9\textwidth}{%
      \caption[]{\label{fig:rap}{\sloppy  (a) Shapes of the \ww{} transverse-momentum distribution
          differential in the rapidity of the \ww{} pair at the \nnll{}\plus\nnlo{} for $|y|<0.5$ (red, solid), 
          $0.5<|y|<1$ (blue, dashed), $1<|y|<2$ (black, dotted), 
          $2<|y|<3$ (magenta, dash-dotted), $3<|y|$ (orange, double-dash dotted); and (b) the shape-ratio with respect to the inclusive result.
}}}
\end{center}
\end{figure}

So far, we only considered \pt{} spectra for on-shell \ww{} and \zz{} production
that are inclusive in the kinematics of 
the vector-boson pair. Our numerical program, however, allows us to compute
arbitrary observables that are differential with respect to the \vvp{} phase space.

In the following we study
the behaviour of the transverse-momentum spectrum in
different rapidity regions of the vector-boson pair. In \fig{fig:rap} we study
the shape of the \nnll{}\plus{}\nnlo{} transverse-momentum 
distribution, i.e.\ normalized such that its integral yields one, for $|y|<0.5$ (red, solid), $0.5<|y|<1$ 
(blue, dashed), $1<|y|<2$ (black, dotted), 
$2<|y|<3$ (magenta, dash-dotted) and $3<|y|$ (orange, dash-double dotted). 
The right panel shows the same results normalized to the fully inclusive distribution.
We clearly see that the 
\pt{} shapes become softer as the rapidity increases. In the central 
region ($|y|<2$) the distributions are still quite insensitive to the specific 
value of the rapidity and only slightly harder than the inclusive spectrum.
In the forward rapidity region, on the other hand, the shapes become increasingly softer.

The observed pattern can be understood in the following way: rapidity 
and transverse momentum are two not completely independent phase-space 
variables. Indeed, they affect their mutual upper integration bounds. At higher 
rapidities the kinematically allowed range of transverse momenta is reduced:
this squeezes the \pt{} spectrum which consequently 
becomes softer. This effect has been observed also in previous studies
in the case of Higgs boson production \cite{Bozzi:2007pn}.

\subsection{The \ww{}  cross section and \pt{}-veto efficiencies}
\label{sec:pteff}

\begin{figure}[h]
\begin{center}
\includegraphics[trim = 10mm -5mm 0mm 0mm, height=.46\textheight]{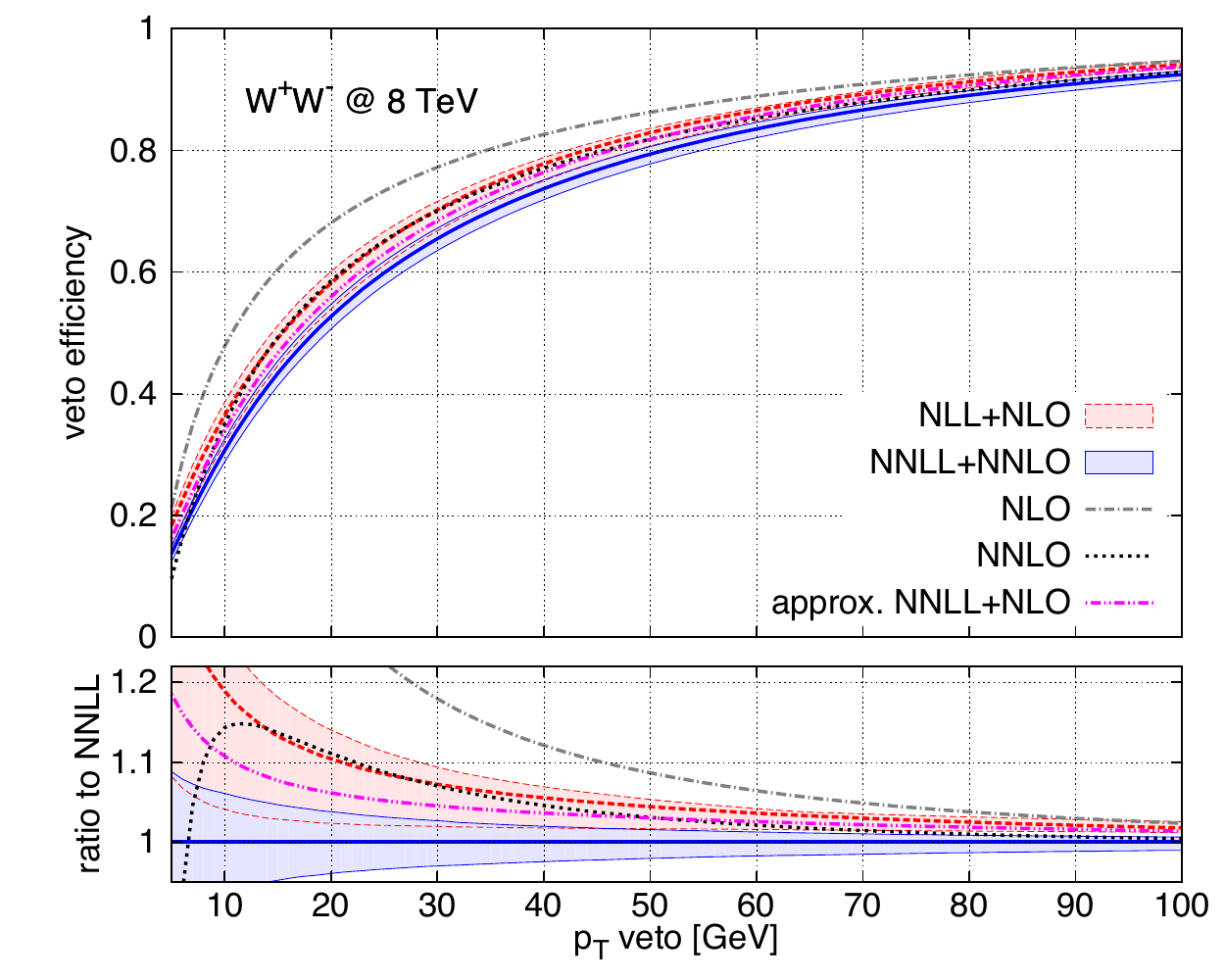}
    \parbox{.9\textwidth}{%
      \caption[]{\label{fig:veto}{\sloppy Veto efficiency for the transverse momentum 
      of the \ww{} pair at various orders: \nll{}\plus\nlo{} (red, dashed), 
      \nnll{}\plus\nnlo{} (blue, solid), \nlo{} (grey, dash-dotted), \nnlo{} (black, dotted),
       approximate \nnll{}\plus{}\nlo{} (magenta, dash-double dotted); thick lines: central scale choices; bands: uncertainty
          due to combined scale variations; thin lines: borders of bands.}  }}
\end{center}
\end{figure}

The excess in the \ww{} production cross section measured by
\atl{} \cite{ATLAS-CONF-2014-033} with respect to the SM prediction
has drawn a lot of attention to the \ww{} process, 
since the \ww{} signature appears in many new physics scenarios \cite{Morrissey:2009tf}.
The inclusion of the recently computed \nnlo{} corrections \cite{Gehrmann:2014fva}
considerably reduces the significance of the excess. However,
particular attention must be payed to
the modelling of the jet veto \cite{Monni:2014zra,Meade:2014fca,CMS:2015uda} 
when extrapolating from the fiducial region
to obtain the inclusive cross section.
Effects of jet veto resummation have been considered in \citeres{Jaiswal:2014yba,Becher:2014aya},
though still matching to the fixed-order ${\cal O}(\as)$ result.

In this paper we are dealing with transverse-momentum spectra, and we perform a resummation
on a different variable with respect to the jet \pt. However, the vector-boson pair \pt{} and
the jet \pt{} are clearly related variables (actually, at ${\cal O}(\as)$ they indeed coincide).
We will therefore study the \pt{}-veto efficiency in \ww{} production
at different orders in resummed and fixed-order perturbation theory.
We define the \pt{}-veto efficiency as
\bal
\epsilon(\pt^{\text{veto}}) = \sigma(\pt<\pt^{\text{veto}}{})/\sigma_\text{tot}\, .
\eal
In \fig{fig:veto} we show $\epsilon(\pt^{\text{veto}})$ at the \nnll{}\plus{}\nnlo{} (blue, solid), 
approximate \nnll{}\plus{}\nlo{} (magenta, dash-double dotted), 
\nll{}\plus{}\nlo{} (red, dashed), \nnlo{} (black, dotted) 
and \nlo{} (grey, dash-dotted). The lower inset shows the same curves 
normalized to our reference prediction at \nnll{}\plus{}\nnlo{}.
Our approximate \nnll{}\plus{}\nlo{} is obtained by simply adding the $g^{(3)}$ function in
the Sudakov exponent in \eqn{exponent} at \nll{}\plus{}\nlo{}, and corresponds to the approximation considered
in \citeres{Grazzini:2005vw,Meade:2014fca}.

For reference, the corresponding numerical values of the efficiencies are 
given in \tab{tab:veto} for $\pt{}=5$-$40$\,GeV. 
The uncertainty bands are obtained by a combined variation of resummation, factorization
and renormalization scales
as in Fig.~\ref{fig:bestpredictionww}.
The first thing we observe is that the \nlo{} result appears to be 
well above the others and cannot be really considered a reliable prediction for the efficiency.
This is because it is essentially a \lo{} prediction at finite values of $\pt^{\text{veto}}$.
We also note that in the small-\pt{} region (say below $\pt{}\sim10$\,GeV)
the fixed-order \nlo{} and \nnlo{} predictions diverge and cannot be trusted.
Comparing further the fixed-order results among 
each other and the resummed results among each 
other, we observe that higher-order corrections in fixed-order and resummed
perturbation theory reduce the \pt{}-veto efficiency. 

Both effects can be easily understood in the light of the results presented up to now.
As seen in \fig{fig:fixedorder}, the inclusion of the \nnlo{} corrections make the \pt{} distribution harder. Furthermore, resummation effects generally harden the spectrum.
A qualitatively similar result is obtained when going from \nll{}\plus{}\nlo{} to \nnll{}\plus{}\nnlo{} (see \fig{fig:bestpredictionww}).

\begin{sloppypar}
\tolerance 9999
It is interesting to compare the approximated \nnll{}\plus{}\nlo{} result with the \nnlo{} and \nnll{}\plus{}\nnlo{} predictions.
For values of $\pt^{\text{veto}}\sim 25-30$ GeV we see that the approximated result is in between the \nnlo{} one
and our best \nnll{}\plus{}\nnlo{} prediction. This means that the effect of \nnll{} resummation obtained by the inclusion
of the $g^{(3)}$ function in the Sudakov exponent in \eqn{exponent} is quantitatively important.
Nonetheless, the efficiency obtained within this approximation is still about $5\%$ higher than the \nnll{}\plus{}\nnlo{} prediction. We also notice that in this region of
$\pt^{\text{veto}}$, the \nnlo{} and \nll{}\plus{}\nlo{} results differ by less than $1\%$.
\end{sloppypar}

Comparing the \nnll{}\plus{}\nnlo{} and \nll{}\plus{}\nlo{} results, we find that
they are compatible within the corresponding uncertainties.


We add few comments on the 
recent measurement of the \ww{} cross section carried out by the \cms{} collaboration \cite{CMS:2015uda}.
The result shows good agreement with the \nnlo{} prediction of \citere{Gehrmann:2014fva}.
The corresponding analysis, however, is based on a reweighting procedure of the \pt{} spectrum of the \ww pair.
The events generated with \powheg{} \cite{Nason:2004rx} 
plus \pythia{6}{} \cite{Sjostrand:2006za} were reweighted by using the calculation of
\citere{Meade:2014fca}, which corresponds to our \nnll{}\plus{}\nlo{} approximation, and includes neither the second-order hard-collinear coefficient ${\cal H}^{WW,(2)}$ in \eqn{hexpan},
nor the \nnlo{} matching.
The results in \fig{fig:veto} show that the \nnll{}\plus{}\nnlo{}
\pt{}-veto efficiency is lower than the efficiency obtained with the approximated \nnll{}\plus{}\nlo{} calculation.
As a consequence, a reweighting to the full
\nnll{}\plus{}\nnlo{} prediction for the \ww{} spectrum
would most likely lead to a decrease of the jet-veto efficiency.


\begin{table}[t]
\begin{center}
\begin{tabular}{c|ccccc}
\toprule
 & 
\multicolumn{4}{c}{$\sigma(\pt<\pt^{\text{veto}}{})/\sigma_\text{tot}$ [\%]} \\
 $\pt^{\text{veto}}$ [GeV] & \nnlo{}\plus{}\nnll{} & approx. \nnll{}\plus{}\nlo{} & \nlo{}\plus{}\nll{} & \nnlo{} & \nlo{}\\
\midrule
5  & 13.7$^{+8.8 \%}_{-9.6\%}$ & 16.2 & 18.3$^{+8.8 \%}_{-19\%}$ & 9.6 & 21.2\Bstrut \\ \Tstrut
10 & 30.7$^{+6.1 \%}_{-6.2\%}$ & 34.0 & 36.5$^{+5.9 \%}_{-12\%}$ & 35.1 & 47.8\Bstrut \\ \Tstrut
15 & 43.4$^{+4.7 \%}_{-4.8\%}$ & 46.7 & 49.1$^{+4.3 \%}_{-9.2\%}$ & 49.3 & 60.4\Bstrut \\ \Tstrut
20 & 52.7$^{+3.8 \%}_{-3.9\%}$ & 55.9 & 58.2$^{+3.3 \%}_{-7.3\%}$ & 58.5 & 68.1\Bstrut \\ \Tstrut
25 & 59.9$^{+3.1 \%}_{-3.4\%}$ & 63.0 & 65.0$^{+2.6 \%}_{-5.9\%}$ & 65.1 & 73.4\Bstrut \\ \Tstrut
30 & 65.5$^{+2.7 \%}_{-3.0\%}$ & 68.4 & 70.2$^{+2.0 \%}_{-4.9\%}$ & 70.1 & 77.2\Bstrut \\ \Tstrut
35 & 70.0$^{+2.3 \%}_{-2.7\%}$ & 72.8 & 74.4$^{+1.6 \%}_{-4.2\%}$ & 74.0 & 80.2\Bstrut \\ \Tstrut
40 & 73.7$^{+2.0 \%}_{-2.5\%}$ & 76.4 & 77.8$^{+1.3 \%}_{-3.6\%}$ & 77.1 & 82.6 \Bstrut \\
\bottomrule
\end{tabular}
\end{center}
\caption{\label{tab:veto}
Predictions for the \pt{}-veto efficiency (in percent) at various perturbative orders.
}
\end{table}

\section{Summary}
\label{sec:summa}

In this paper we have studied the transverse-momentum distribution of vector-boson pairs
in hadronic collisions. We presented a computation of the \pt{} spectrum in which
the logarithmically enhanced contributions at small \pt{} are resummed up to \nnll{} accuracy
and the ensuing result is combined
with state-of-the-art ${\cal O}(\as^2)$ (\nnlo{}) predictions valid at large \pt{}.
We presented numerical results for \ww{} and \zz{} production at the LHC together with
a study of their perturbative uncertainties.

We found that, up to relatively large transverse momenta,
when scale variations are studied around
the fixed resummation scale $\Qres=m_V$ 
we obtain results which are fully consistent with those
obtained using the dynamical choice $\Qres=M_{VV}/2$.
At large transverse momenta
the fixed-order result in the tail of the distribution is nicely recovered with the 
fixed-scale choice.
Our new \nnll{}\plus{}\nnlo{} results
significantly reduce the theoretical uncertainties obtained through scale 
variations compared to lower orders in both the peak region and the tail of the 
distribution.

We have also studied
the rapidity dependence of the resummed transverse-momentum distribution.
The rapidity dependence at \nnll{}\plus{}\nnlo{} is quite flat
in the central region ($|y|\lesssim 2$), but signals a substantially 
softer spectrum in the forward region. Due to phase-space suppression, the effect 
on the inclusive transverse-momentum distribution is very moderate though.

Finally, we have studied the \pt{}-veto efficiency at different orders in
resummed and fixed-order perturbation theory. 
Both \nnll{} resummation and the \nnlo{} effects turned out to be
important to obtain an accurate prediction for this quantity. We observed that the veto efficiency at \nnll{}\plus{}\nnlo{} is
$\sim 5$\% lower ($\sim 3\%$ in absolute terms)
with respect to the approximate \nnll{}\plus{}\nlo{} calculation
used in the \cms{} analysis of \citere{CMS:2015uda}.
This result suggests that our \nnll{}\plus{}\nnlo{} predictions
will be useful to validate the transverse-momentum spectra obtained
from Monte Carlo event generators,
similarly to what was done for the \nnll{}\plus{}\nnlo{} calculation of \citere{Bozzi:2005wk}
in the case of Higgs boson production.

In this paper we considered the \pt{} spectrum of stable vector-boson pairs.
Exploiting the two-loop helicity amplitudes
for $q{\bar q}\to VV^\prime\to 4$ leptons \cite{Caola:2014iua,Gehrmann:2015ora} will allow us to extend the
calculation to include the leptonic decay of the vector bosons and off-shell effects. 
The computation of the transverse-momentum spectrum
with realistic experimental cuts will then become possible.

\noindent {\bf Acknowledgements.}
We would like to thank Andreas von Manteuffel and Lorenzo Tancredi for providing 
us with their private code to evaluate the helicity-averaged on-shell $VV'$ amplitudes in the equal-mass case. We would like to thank Giancarlo Ferrera for comments on the manuscript. This research was supported in part by the Swiss National Science Foundation ({\abbrev SNF}) under contracts CRSII2-141847, 200021-156585 and by the Research Executive Agency (REA) of the 
European Union under the Grant Agreement number PITN-GA-2012-316704 ({\it Higgstools}).


\renewcommand{\em}{}
\bibliographystyle{UTPstyle}
\bibliography{final}

\providecommand{\href}[2]{#2}\begingroup\raggedright\begin{thebibliography}{10}

\bibitem{Aad:2012tfa}
{\bf ATLAS} Collaboration, {\it {Observation of a new particle in the search
  for the Standard Model Higgs boson with the ATLAS detector at the LHC}},
  {\em Phys.Lett.} {\bf B716} (2012) 1--29,
  [\href{http://xxx.lanl.gov/abs/1207.7214}{{\tt arXiv:1207.7214}}].

\bibitem{Chatrchyan:2012ufa}
{\bf CMS} Collaboration, {\it {Observation of a new boson at a mass of 125 GeV
  with the CMS experiment at the LHC}},  {\em Phys.Lett.} {\bf B716} (2012)
  30--61, [\href{http://xxx.lanl.gov/abs/1207.7235}{{\tt arXiv:1207.7235}}].

\bibitem{Wang:2014uea}
{\bf ATLAS, D0, CDF, CMS} Collaboration, {\it {Diboson Production at LHC and
  Tevatron}},  {\em Int.J.Mod.Phys.Conf.Ser.} {\bf 31} (2014) 1460279,
  [\href{http://xxx.lanl.gov/abs/1403.1415}{{\tt arXiv:1403.1415}}].

\bibitem{Kauer:2012hd}
N.~Kauer and G.~Passarino, {\it {Inadequacy of zero-width approximation for a
  light Higgs boson signal}},  {\em JHEP} {\bf 1208} (2012) 116,
  [\href{http://xxx.lanl.gov/abs/1206.4803}{{\tt arXiv:1206.4803}}].

\bibitem{Caola:2013yja}
F.~Caola and K.~Melnikov, {\it {Constraining the Higgs boson width with $ZZ$
  production at the LHC}},  {\em Phys.Rev.} {\bf D88} (2013) 054024,
  [\href{http://xxx.lanl.gov/abs/1307.4935}{{\tt arXiv:1307.4935}}].

\bibitem{Campbell:2013wga}
J.~M. Campbell, R.~K. Ellis, and C.~Williams, {\it {Bounding the Higgs width at
  the LHC: Complementary results from $H \to WW$}},  {\em Phys.Rev.} {\bf D89}
  (2014), no.~5 053011, [\href{http://xxx.lanl.gov/abs/1312.1628}{{\tt
  arXiv:1312.1628}}].

\bibitem{Ohnemus:1990za}
J.~Ohnemus and J.~Owens, {\it {An Order $\alpha_s$ calculation of hadronic $Z
  Z$ production}},  {\em Phys.Rev.} {\bf D43} (1991) 3626--3639.

\bibitem{Mele:1990bq}
B.~Mele, P.~Nason, and G.~Ridolfi, {\it {QCD radiative corrections to Z boson
  pair production in hadronic collisions}},  {\em Nucl.Phys.} {\bf B357} (1991)
  409--438.

\bibitem{Ohnemus:1994ff}
J.~Ohnemus, {\it {Hadronic $Z Z$, $W^{-} W^{+}$, and $W^\pm Z$ production with
  QCD corrections and leptonic decays}},  {\em Phys.Rev.} {\bf D50} (1994)
  1931--1945, [\href{http://xxx.lanl.gov/abs/hep-ph/9403331}{{\tt
  hep-ph/9403331}}].

\bibitem{Campbell:1999ah}
J.~M. Campbell and R.~K. Ellis, {\it {An Update on vector boson pair production
  at hadron colliders}},  {\em Phys.Rev.} {\bf D60} (1999) 113006,
  [\href{http://xxx.lanl.gov/abs/hep-ph/9905386}{{\tt hep-ph/9905386}}].

\bibitem{Dixon:1999di}
L.~J. Dixon, Z.~Kunszt, and A.~Signer, {\it {Vector boson pair production in
  hadronic collisions at order $\alpha_s$ : Lepton correlations and anomalous
  couplings}},  {\em Phys.Rev.} {\bf D60} (1999) 114037,
  [\href{http://xxx.lanl.gov/abs/hep-ph/9907305}{{\tt hep-ph/9907305}}].

\bibitem{Dixon:1998py}
L.~J. Dixon, Z.~Kunszt, and A.~Signer, {\it {Helicity amplitudes for
  O($\alpha_s$) production of $W^{+} W^{-}$, $W^\pm Z$, $Z Z$, $W^\pm \gamma$,
  or $Z \gamma$ pairs at hadron colliders}},  {\em Nucl.Phys.} {\bf B531}
  (1998) 3--23, [\href{http://xxx.lanl.gov/abs/hep-ph/9803250}{{\tt
  hep-ph/9803250}}].

\bibitem{Dicus:1987dj}
D.~A. Dicus, C.~Kao, and W.~Repko, {\it {Gluon Production of Gauge Bosons}},
  {\em Phys.Rev.} {\bf D36} (1987) 1570.

\bibitem{vanderBij:1988fb}
J.~van~der Bij and E.~N. Glover, {\it {Photon $Z$ boson pair production via
  gluon fusion}},  {\em Phys.Lett.} {\bf B206} (1988) 701.

\bibitem{Matsuura:1991pj}
T.~Matsuura and J.~van~der Bij, {\it {Characteristics of leptonic signals for Z
  boson pairs at hadron colliders}},  {\em Z.Phys.} {\bf C51} (1991) 259--266.

\bibitem{Zecher:1994kb}
C.~Zecher, T.~Matsuura, and J.~van~der Bij, {\it {Leptonic signals from
  off-shell Z boson pairs at hadron colliders}},  {\em Z.Phys.} {\bf C64}
  (1994) 219--226, [\href{http://xxx.lanl.gov/abs/hep-ph/9404295}{{\tt
  hep-ph/9404295}}].

\bibitem{Binoth:2008pr}
T.~Binoth, N.~Kauer, and P.~Mertsch, {\it {Gluon-induced QCD corrections to $pp
  \rightarrow ZZ \rightarrow l\bar{l} l^\prime \bar{l}^\prime$}},
  \href{http://xxx.lanl.gov/abs/0807.0024}{{\tt arXiv:0807.0024}}.

\bibitem{Campbell:2011bn}
J.~M. Campbell, R.~K. Ellis, and C.~Williams, {\it {Vector boson pair
  production at the LHC}},  {\em JHEP} {\bf 1107} (2011) 018,
  [\href{http://xxx.lanl.gov/abs/1105.0020}{{\tt arXiv:1105.0020}}].

\bibitem{Bierweiler:2013dja}
A.~Bierweiler, T.~Kasprzik, and J.~H. Kühn, {\it {Vector-boson pair production
  at the LHC to $\mathcal{O}(\alpha^3)$ accuracy}},  {\em JHEP} {\bf 1312}
  (2013) 071, [\href{http://xxx.lanl.gov/abs/1305.5402}{{\tt
  arXiv:1305.5402}}].

\bibitem{Baglio:2013toa}
J.~Baglio, L.~D. Ninh, and M.~M. Weber, {\it {Massive gauge boson pair
  production at the LHC: a next-to-leading order story}},  {\em Phys.Rev.} {\bf
  D88} (2013) 113005, [\href{http://xxx.lanl.gov/abs/1307.4331}{{\tt
  arXiv:1307.4331}}].

\bibitem{Binoth:2009wk}
T.~Binoth, T.~Gleisberg, S.~Karg, N.~Kauer, and G.~Sanguinetti, {\it {NLO QCD
  corrections to $ZZ$+jet production at hadron colliders}},  {\em Phys.Lett.}
  {\bf B683} (2010) 154--159, [\href{http://xxx.lanl.gov/abs/0911.3181}{{\tt
  arXiv:0911.3181}}].

\bibitem{Binoth:2010nha}
{\bf SM and NLO Multileg Working Group} Collaboration, T.~Binoth et~al., {\it
  {The SM and NLO Multileg Working Group: Summary report}},
  \href{http://xxx.lanl.gov/abs/1003.1241}{{\tt arXiv:1003.1241}}.

\bibitem{Cascioli:2014yka}
F.~Cascioli, T.~Gehrmann, M.~Grazzini, S.~Kallweit, P.~Maierh\"ofer, A.~von
  Manteuffel, S.~Pozzorini, D.~Rathlev, L.~Tancredi, and E.~Weihs, {\it {$ZZ$
  production at hadron colliders in NNLO QCD}},  {\em Phys.Lett.} {\bf B735}
  (2014) 311--313, [\href{http://xxx.lanl.gov/abs/1405.2219}{{\tt
  arXiv:1405.2219}}].

\bibitem{Ohnemus:1991kk}
J.~Ohnemus, {\it {An Order $\alpha_s$ calculation of hadronic $W^{-} W^{+}$
  production}},  {\em Phys.Rev.} {\bf D44} (1991) 1403--1414.

\bibitem{Frixione:1993yp}
S.~Frixione, {\it {A Next-to-leading order calculation of the cross-section for
  the production of $W^+ W^-$ pairs in hadronic collisions}},  {\em Nucl.Phys.}
  {\bf B410} (1993) 280--324.

\bibitem{Glover:1988fe}
E.~N. Glover and J.~van~der Bij, {\it {Vector boson pair production via gluon
  fusion}},  {\em Phys.Lett.} {\bf B219} (1989) 488.

\bibitem{Binoth:2005ua}
T.~Binoth, M.~Ciccolini, N.~Kauer, and M.~Kr\"amer, {\it {Gluon-induced $WW$
  background to Higgs boson searches at the LHC}},  {\em JHEP} {\bf 0503}
  (2005) 065, [\href{http://xxx.lanl.gov/abs/hep-ph/0503094}{{\tt
  hep-ph/0503094}}].

\bibitem{Binoth:2006mf}
T.~Binoth, M.~Ciccolini, N.~Kauer, and M.~Kr\"amer, {\it {Gluon-induced W-boson
  pair production at the LHC}},  {\em JHEP} {\bf 0612} (2006) 046,
  [\href{http://xxx.lanl.gov/abs/hep-ph/0611170}{{\tt hep-ph/0611170}}].

\bibitem{Campbell:2011cu}
J.~M. Campbell, R.~K. Ellis, and C.~Williams, {\it {Gluon-Gluon Contributions
  to $W^+ W^-$ Production and Higgs Interference Effects}},  {\em JHEP} {\bf
  1110} (2011) 005, [\href{http://xxx.lanl.gov/abs/1107.5569}{{\tt
  arXiv:1107.5569}}].

\bibitem{Bierweiler:2012kw}
A.~Bierweiler, T.~Kasprzik, J.~H. Kühn, and S.~Uccirati, {\it {Electroweak
  corrections to W-boson pair production at the LHC}},  {\em JHEP} {\bf 1211}
  (2012) 093, [\href{http://xxx.lanl.gov/abs/1208.3147}{{\tt
  arXiv:1208.3147}}].

\bibitem{Billoni:2013aba}
M.~Billoni, S.~Dittmaier, B.~Jäger, and C.~Speckner, {\it {Next-to-leading
  order electroweak corrections to $pp\rightarrow W^+W^- \rightarrow 4$ leptons
  at the LHC in double-pole approximation}},  {\em JHEP} {\bf 1312} (2013) 043,
  [\href{http://xxx.lanl.gov/abs/1310.1564}{{\tt arXiv:1310.1564}}].

\bibitem{Dittmaier:2007th}
S.~Dittmaier, S.~Kallweit, and P.~Uwer, {\it {NLO QCD corrections to $WW$+jet
  production at hadron colliders}},  {\em Phys.Rev.Lett.} {\bf 100} (2008)
  062003, [\href{http://xxx.lanl.gov/abs/0710.1577}{{\tt arXiv:0710.1577}}].

\bibitem{Campbell:2007ev}
J.~M. Campbell, R.~K. Ellis, and G.~Zanderighi, {\it {Next-to-leading order
  predictions for $WW+1$ jet distributions at the LHC}},  {\em JHEP} {\bf 0712}
  (2007) 056, [\href{http://xxx.lanl.gov/abs/0710.1832}{{\tt
  arXiv:0710.1832}}].

\bibitem{Dittmaier:2009un}
S.~Dittmaier, S.~Kallweit, and P.~Uwer, {\it {NLO QCD corrections to
  $pp/p\bar{p}\rightarrow WW$+jet+X including leptonic $W$-boson decays}},
  {\em Nucl.Phys.} {\bf B826} (2010) 18--70,
  [\href{http://xxx.lanl.gov/abs/0908.4124}{{\tt arXiv:0908.4124}}].

\bibitem{Cascioli:2013gfa}
F.~Cascioli, S.~H\"oche, F.~Krauss, P.~Maierh\"ofer, S.~Pozzorini, and
  F.~Siegert, {\it {Precise Higgs-background predictions: merging NLO QCD and
  squared quark-loop corrections to four-lepton + 0,1 jet production}},  {\em
  JHEP} {\bf 1401} (2014) 046, [\href{http://xxx.lanl.gov/abs/1309.0500}{{\tt
  arXiv:1309.0500}}].

\bibitem{Gehrmann:2014fva}
T.~Gehrmann, M.~Grazzini, S.~Kallweit, , P.~Maierh\"ofer, A.~von Manteuffel,
  S.~Pozzorini, D.~Rathlev, and L.~Tancredi, {\it {$W^+W^-$ Production at
  Hadron Colliders in Next to Next to Leading Order QCD}},  {\em
  Phys.Rev.Lett.} {\bf 113} (2014) 212001,
  [\href{http://xxx.lanl.gov/abs/1408.5243}{{\tt arXiv:1408.5243}}].

\bibitem{Caola:2014iua}
F.~Caola, J.~M. Henn, K.~Melnikov, A.~V. Smirnov, and V.~A. Smirnov, {\it
  {Two-loop helicity amplitudes for the production of two off-shell electroweak
  bosons in quark-antiquark collisions}},  {\em JHEP} {\bf 1411} (2014) 041,
  [\href{http://xxx.lanl.gov/abs/1408.6409}{{\tt arXiv:1408.6409}}].

\bibitem{Gehrmann:2015ora}
T.~Gehrmann, A.~von Manteuffel, and L.~Tancredi, {\it {The two-loop helicity
  amplitudes for $q \bar q' \to V_1 V_2 \to 4~\mathrm{leptons}$}},
  \href{http://xxx.lanl.gov/abs/1503.04812}{{\tt arXiv:1503.04812}}.

\bibitem{Caola:2015ila}
F.~Caola, J.~M. Henn, K.~Melnikov, A.~V. Smirnov, and V.~A. Smirnov, {\it
  {Two-loop helicity amplitudes for the production of two off-shell electroweak
  bosons in gluon fusion}},  {\em JHEP} {\bf 1506} (2015) 129,
  [\href{http://xxx.lanl.gov/abs/1503.08759}{{\tt arXiv:1503.08759}}].

\bibitem{vonManteuffel:2015msa}
A.~von Manteuffel and L.~Tancredi, {\it {The two-loop helicity amplitudes for
  $gg \to V_1 V_2 \to 4~\mathrm{leptons}$}},  {\em JHEP} {\bf 1506} (2015) 197,
  [\href{http://xxx.lanl.gov/abs/1503.08835}{{\tt arXiv:1503.08835}}].

\bibitem{CDF:2011ab}
{\bf CDF} Collaboration, {\it {Measurement of $ZZ$ production in leptonic final
  states at $\sqrt{s}$ of 1.96 TeV at CDF}},  {\em Phys.Rev.Lett.} {\bf 108}
  (2012) 101801, [\href{http://xxx.lanl.gov/abs/1112.2978}{{\tt
  arXiv:1112.2978}}].

\bibitem{Abazov:2012cj}
{\bf D0} Collaboration, {\it {A measurement of the $WZ$ and $ZZ$ production
  cross sections using leptonic final states in 8.6 fb$^{-1}$ of $p\bar{p}$
  collisions}},  {\em Phys.Rev.} {\bf D85} (2012) 112005,
  [\href{http://xxx.lanl.gov/abs/1201.5652}{{\tt arXiv:1201.5652}}].

\bibitem{Aad:2012awa}
{\bf ATLAS} Collaboration, {\it {Measurement of $ZZ$ production in $pp$
  collisions at $\sqrt{s}=7$ TeV and limits on anomalous $ZZZ$ and $ZZ\gamma$
  couplings with the ATLAS detector}},  {\em JHEP} {\bf 1303} (2013) 128,
  [\href{http://xxx.lanl.gov/abs/1211.6096}{{\tt arXiv:1211.6096}}].

\bibitem{Chatrchyan:2012sga}
{\bf CMS} Collaboration, {\it {Measurement of the $ZZ$ production cross section
  and search for anomalous couplings in 2l2l' final states in $pp$ collisions
  at $\sqrt{s}=7$ TeV}},  {\em JHEP} {\bf 1301} (2013) 063,
  [\href{http://xxx.lanl.gov/abs/1211.4890}{{\tt arXiv:1211.4890}}].

\bibitem{ATLAS:2013gma}
{\bf ATLAS} Collaboration, {\it {Measurement of the total $ZZ$ production cross
  section in proton-proton collisions at $\sqrt{s} =8$ in 20 fb$^{-1}$ with the
  ATLAS detector}},  \href{http://xxx.lanl.gov/abs/ATLAS-CONF-2013-020,
  ATLAS-COM-CONF-2013-020}{{\tt ATLAS-CONF-2013-020, ATLAS-COM-CONF-2013-020}}.

\bibitem{Chatrchyan:2013oev}
{\bf CMS} Collaboration, {\it {Measurement of $W^+ W^-$ and $ZZ$ production
  cross sections in pp collisions at $\sqrt{s} = 8$\,TeV}},  {\em Phys.Lett.}
  {\bf B721} (2013) 190--211, [\href{http://xxx.lanl.gov/abs/1301.4698}{{\tt
  arXiv:1301.4698}}].

\bibitem{CMS:2014xja}
{\bf CMS} Collaboration, {\it {Measurement of the $pp \to ZZ$ production cross
  section and constraints on anomalous triple gauge couplings in four-lepton
  final states at $\sqrt s=$8 TeV}},  {\em Phys.Lett.} {\bf B740} (2015)
  250--272, [\href{http://xxx.lanl.gov/abs/1406.0113}{{\tt arXiv:1406.0113}}].

\bibitem{Abazov:2011cb}
{\bf D0} Collaboration, {\it {Measurements of $WW$ and $WZ$ production in $W$+
  jets final states in $p\bar{p}$ collisions}},  {\em Phys.Rev.Lett.} {\bf 108}
  (2012) 181803, [\href{http://xxx.lanl.gov/abs/1112.0536}{{\tt
  arXiv:1112.0536}}].

\bibitem{ATLAS:2012mec}
{\bf ATLAS} Collaboration, {\it {Measurement of $W^+W^-$ production in $pp$
  collisions at $\sqrt{s}$=7\,TeV with the ATLAS detector and limits on
  anomalous $WWZ$ and $WW\gamma$ couplings}},  {\em Phys.Rev.} {\bf D87}
  (2013), no.~11 112001, [\href{http://xxx.lanl.gov/abs/1210.2979}{{\tt
  arXiv:1210.2979}}].

\bibitem{Chatrchyan:2013yaa}
{\bf CMS} Collaboration, {\it {Measurement of the $W^+W^-$ Cross section in
  $pp$ Collisions at $\sqrt{s} = 7$ TeV and Limits on Anomalous $WW\gamma$ and
  $WWZ$ couplings}},  {\em Eur.Phys.J.} {\bf C73} (2013), no.~10 2610,
  [\href{http://xxx.lanl.gov/abs/1306.1126}{{\tt arXiv:1306.1126}}].

\bibitem{ATLAS-CONF-2014-033}
{\bf ATLAS} Collaboration, {\it {Measurement of the $W^+W^-$ production cross
  section in proton-proton collisions at $\sqrt{s} =8$ TeV with the ATLAS
  detector}},  \href{http://xxx.lanl.gov/abs/ATLAS-CONF-2014-033,
  ATLAS-COM-CONF-2014-045}{{\tt ATLAS-CONF-2014-033, ATLAS-COM-CONF-2014-045}}.

\bibitem{CMS:2015uda}
{\bf CMS} Collaboration, {\it {Measurement of the $W^+W^-$ cross section in pp
  collisions at sqrt(s) = 8 TeV and limits on anomalous gauge couplings}},
  \href{http://xxx.lanl.gov/abs/CMS-PAS-SMP-14-016}{{\tt CMS-PAS-SMP-14-016}}.

\bibitem{Morrissey:2009tf}
D.~E. Morrissey, T.~Plehn, and T.~M. Tait, {\it {Physics searches at the LHC}},
   {\em Phys.Rept.} {\bf 515} (2012) 1--113,
  [\href{http://xxx.lanl.gov/abs/0912.3259}{{\tt arXiv:0912.3259}}].

\bibitem{Balazs:1998bm}
C.~Balazs and C.~Yuan, {\it {Higgs boson production at hadron colliders with
  soft gluon effects. 1. Backgrounds}},  {\em Phys.Rev.} {\bf D59} (1999)
  114007, [\href{http://xxx.lanl.gov/abs/hep-ph/9810319}{{\tt
  hep-ph/9810319}}].

\bibitem{Grazzini:2005vw}
M.~Grazzini, {\it {Soft-gluon effects in $WW$ production at hadron colliders}},
   {\em JHEP} {\bf 0601} (2006) 095,
  [\href{http://xxx.lanl.gov/abs/hep-ph/0510337}{{\tt hep-ph/0510337}}].

\bibitem{Frederix:2008vb}
R.~Frederix and M.~Grazzini, {\it {Higher-order QCD effects in the $h
  \rightarrow ZZ$ search channel at the LHC}},  {\em Phys.Lett.} {\bf B662}
  (2008) 353--359, [\href{http://xxx.lanl.gov/abs/0801.2229}{{\tt
  arXiv:0801.2229}}].

\bibitem{Wang:2013qua}
Y.~Wang, C.~S. Li, Z.~L. Liu, D.~Y. Shao, and H.~T. Li, {\it
  {Transverse-Momentum Resummation for Gauge Boson Pair Production at the
  Hadron Collider}},  {\em Phys.Rev.} {\bf D88} (2013) 114017,
  [\href{http://xxx.lanl.gov/abs/1307.7520}{{\tt arXiv:1307.7520}}].

\bibitem{Meade:2014fca}
P.~Meade, H.~Ramani, and M.~Zeng, {\it {Transverse momentum resummation effects
  in $W^+W^-$ measurements}},  {\em Phys.Rev.} {\bf D90} (2014), no.~11 114006,
  [\href{http://xxx.lanl.gov/abs/1407.4481}{{\tt arXiv:1407.4481}}].

\bibitem{Bozzi:2005wk}
G.~Bozzi, S.~Catani, D.~de~Florian, and M.~Grazzini, {\it {Transverse-momentum
  resummation and the spectrum of the Higgs boson at the LHC}},  {\em
  Nucl.Phys.} {\bf B737} (2006) 73--120,
  [\href{http://xxx.lanl.gov/abs/hep-ph/0508068}{{\tt hep-ph/0508068}}].

\bibitem{deFlorian:2012mx}
D.~de~Florian, G.~Ferrera, M.~Grazzini, and D.~Tommasini, {\it {Higgs boson
  production at the LHC: transverse momentum resummation effects in the
  $H\rightarrow \gamma\gamma$, $H\rightarrow WW \rightarrow l\nu l\nu$ and
  $H\rightarrow ZZ\rightarrow 4l$ decay modes}},  {\em JHEP} {\bf 1206} (2012)
  132, [\href{http://xxx.lanl.gov/abs/1203.6321}{{\tt arXiv:1203.6321}}].

\bibitem{Catani:2015vma}
S.~Catani, D.~de~Florian, G.~Ferrera, and M.~Grazzini, {\it {Vector boson
  production at hadron colliders: transverse-momentum resummation and leptonic
  decay}},  \href{http://xxx.lanl.gov/abs/1507.06937}{{\tt arXiv:1507.06937}}.

\bibitem{Cieri:2015rqa}
L.~Cieri, F.~Coradeschi, and D.~de~Florian, {\it {Diphoton production at hadron
  colliders: transverse-momentum resummation at next-to-next-to-leading
  logarithmic accuracy}},  \href{http://xxx.lanl.gov/abs/1505.03162}{{\tt
  arXiv:1505.03162}}.

\bibitem{Catani:2007vq}
S.~Catani and M.~Grazzini, {\it {An NNLO subtraction formalism in hadron
  collisions and its application to Higgs boson production at the LHC}},  {\em
  Phys.Rev.Lett.} {\bf 98} (2007) 222002,
  [\href{http://xxx.lanl.gov/abs/hep-ph/0703012}{{\tt hep-ph/0703012}}].

\bibitem{Bozzi:2007pn}
G.~Bozzi, S.~Catani, D.~de~Florian, and M.~Grazzini, {\it {Higgs boson
  production at the LHC: Transverse-momentum resummation and rapidity
  dependence}},  {\em Nucl.Phys.} {\bf B791} (2008) 1--19,
  [\href{http://xxx.lanl.gov/abs/0705.3887}{{\tt arXiv:0705.3887}}].

\bibitem{Catani:2013tia}
S.~Catani, L.~Cieri, D.~de~Florian, G.~Ferrera, and M.~Grazzini, {\it
  {Universality of transverse-momentum resummation and hard factors at the
  NNLO}},  {\em Nucl.Phys.} {\bf B881} (2014) 414--443,
  [\href{http://xxx.lanl.gov/abs/1311.1654}{{\tt arXiv:1311.1654}}].

\bibitem{Parisi:1979se}
G.~Parisi and R.~Petronzio, {\it {Small Transverse Momentum Distributions in
  Hard Processes}},  {\em Nucl.Phys.} {\bf B154} (1979) 427.

\bibitem{Curci:1979bg}
G.~Curci, M.~Greco, and Y.~Srivastava, {\it {{QCD} Jets From Coherent States}},
   {\em Nucl.Phys.} {\bf B159} (1979) 451.

\bibitem{Collins:1984kg}
J.~C. Collins, D.~E. Soper, and G.~F. Sterman, {\it {Transverse Momentum
  Distribution in Drell-Yan Pair and W and Z Boson Production}},  {\em
  Nucl.Phys.} {\bf B250} (1985) 199.

\bibitem{Kodaira:1981nh}
J.~Kodaira and L.~Trentadue, {\it {Summing Soft Emission in QCD}},  {\em
  Phys.Lett.} {\bf B112} (1982) 66.

\bibitem{Becher:2010tm}
T.~Becher and M.~Neubert, {\it {{Drell-Yan} Production at Small $q_T$,
  Transverse Parton Distributions and the Collinear Anomaly}},  {\em
  Eur.Phys.J.} {\bf C71} (2011) 1665,
  [\href{http://xxx.lanl.gov/abs/1007.4005}{{\tt arXiv:1007.4005}}].

\bibitem{Davies:1984hs}
C.~Davies and W.~J. Stirling, {\it {Nonleading Corrections to the Drell-Yan
  Cross-Section at Small Transverse Momentum}},  {\em Nucl.Phys.} {\bf B244}
  (1984) 337.

\bibitem{deFlorian:2000pr}
D.~de~Florian and M.~Grazzini, {\it {Next-to-next-to-leading logarithmic
  corrections at small transverse momentum in hadronic collisions}},  {\em
  Phys.Rev.Lett.} {\bf 85} (2000) 4678--4681,
  [\href{http://xxx.lanl.gov/abs/hep-ph/0008152}{{\tt hep-ph/0008152}}].

\bibitem{deFlorian:2001zd}
D.~de~Florian and M.~Grazzini, {\it {The Structure of large logarithmic
  corrections at small transverse momentum in hadronic collisions}},  {\em
  Nucl.Phys.} {\bf B616} (2001) 247--285,
  [\href{http://xxx.lanl.gov/abs/hep-ph/0108273}{{\tt hep-ph/0108273}}].

\bibitem{Catani:2011kr}
S.~Catani and M.~Grazzini, {\it {Higgs Boson Production at Hadron Colliders:
  Hard-Collinear Coefficients at the NNLO}},  {\em Eur.Phys.J.} {\bf C72}
  (2012) 2013, [\href{http://xxx.lanl.gov/abs/1106.4652}{{\tt
  arXiv:1106.4652}}].

\bibitem{Catani:2012qa}
S.~Catani, L.~Cieri, D.~de~Florian, G.~Ferrera, and M.~Grazzini, {\it {Vector
  boson production at hadron colliders: hard-collinear coefficients at the
  NNLO}},  {\em Eur.Phys.J.} {\bf C72} (2012) 2195,
  [\href{http://xxx.lanl.gov/abs/1209.0158}{{\tt arXiv:1209.0158}}].

\bibitem{Grazzini:2013bna}
M.~Grazzini, S.~Kallweit, D.~Rathlev, and A.~Torre, {\it {$Z\gamma$ production
  at hadron colliders in NNLO QCD}},  {\em Phys.Lett.} {\bf B731} (2014)
  204--207, [\href{http://xxx.lanl.gov/abs/1309.7000}{{\tt arXiv:1309.7000}}].

\bibitem{Grazzini:2015nwa}
M.~Grazzini, S.~Kallweit, and D.~Rathlev, {\it {$W\gamma$ and $Z\gamma$
  production at the LHC in NNLO QCD}},
  \href{http://xxx.lanl.gov/abs/1504.01330}{{\tt arXiv:1504.01330}}.

\bibitem{Gehrmann:2012ze}
T.~Gehrmann, T.~L\"ubbert, and L.~L. Yang, {\it {Transverse parton distribution
  functions at next-to-next-to-leading order: the quark-to-quark case}},  {\em
  Phys.Rev.Lett.} {\bf 109} (2012) 242003,
  [\href{http://xxx.lanl.gov/abs/1209.0682}{{\tt arXiv:1209.0682}}].

\bibitem{Gehrmann:2014yya}
T.~Gehrmann, T.~L\"ubbert, and L.~L. Yang, {\it {Calculation of the transverse
  parton distribution functions at next-to-next-to-leading order}},  {\em JHEP}
  {\bf 1406} (2014) 155, [\href{http://xxx.lanl.gov/abs/1403.6451}{{\tt
  arXiv:1403.6451}}].

\bibitem{Catani:1996jh}
S.~Catani and M.~Seymour, {\it {The Dipole formalism for the calculation of QCD
  jet cross-sections at next-to-leading order}},  {\em Phys.Lett.} {\bf B378}
  (1996) 287--301, [\href{http://xxx.lanl.gov/abs/hep-ph/9602277}{{\tt
  hep-ph/9602277}}].

\bibitem{Catani:1996vz}
S.~Catani and M.~Seymour, {\it {A General algorithm for calculating jet
  cross-sections in NLO QCD}},  {\em Nucl.Phys.} {\bf B485} (1997) 291--419,
  [\href{http://xxx.lanl.gov/abs/hep-ph/9605323}{{\tt hep-ph/9605323}}].

\bibitem{Cascioli:2011va}
F.~Cascioli, P.~Maierh\"ofer, and S.~Pozzorini, {\it {Scattering Amplitudes
  with Open Loops}},  {\em Phys.Rev.Lett.} {\bf 108} (2012) 111601,
  [\href{http://xxx.lanl.gov/abs/1111.5206}{{\tt arXiv:1111.5206}}].

\bibitem{Denner:2014gla}
A.~Denner, S.~Dittmaier, and L.~Hofer, {\it {COLLIER - A fortran-library for
  one-loop integrals}},  {\em PoS} {\bf LL2014} (2014) 071,
  [\href{http://xxx.lanl.gov/abs/1407.0087}{{\tt arXiv:1407.0087}}].

\bibitem{Denner:2002ii}
A.~Denner and S.~Dittmaier, {\it {Reduction of one loop tensor five point
  integrals}},  {\em Nucl.Phys.} {\bf B658} (2003) 175--202,
  [\href{http://xxx.lanl.gov/abs/hep-ph/0212259}{{\tt hep-ph/0212259}}].

\bibitem{Denner:2005nn}
A.~Denner and S.~Dittmaier, {\it {Reduction schemes for one-loop tensor
  integrals}},  {\em Nucl.Phys.} {\bf B734} (2006) 62--115,
  [\href{http://xxx.lanl.gov/abs/hep-ph/0509141}{{\tt hep-ph/0509141}}].

\bibitem{Denner:2010tr}
A.~Denner and S.~Dittmaier, {\it {Scalar one-loop 4-point integrals}},  {\em
  Nucl.Phys.} {\bf B844} (2011) 199--242,
  [\href{http://xxx.lanl.gov/abs/1005.2076}{{\tt arXiv:1005.2076}}].

\bibitem{Ossola:2007ax}
G.~Ossola, C.~G. Papadopoulos, and R.~Pittau, {\it {CutTools: A Program
  implementing the OPP reduction method to compute one-loop amplitudes}},  {\em
  JHEP} {\bf 0803} (2008) 042, [\href{http://xxx.lanl.gov/abs/0711.3596}{{\tt
  arXiv:0711.3596}}].

\bibitem{vanHameren:2010cp}
A.~van Hameren, {\it {OneLOop: For the evaluation of one-loop scalar
  functions}},  {\em Comput.Phys.Commun.} {\bf 182} (2011) 2427--2438,
  [\href{http://xxx.lanl.gov/abs/1007.4716}{{\tt arXiv:1007.4716}}].

\bibitem{Ball:2014uwa}
{\bf NNPDF} Collaboration, R.~D. Ball et~al., {\it {Parton distributions for
  the LHC Run II}},  {\em JHEP} {\bf 1504} (2015) 040,
  [\href{http://xxx.lanl.gov/abs/1410.8849}{{\tt arXiv:1410.8849}}].

\bibitem{Bozzi:2010xn}
G.~Bozzi, S.~Catani, G.~Ferrera, D.~de~Florian, and M.~Grazzini, {\it
  {Production of Drell-Yan lepton pairs in hadron collisions:
  Transverse-momentum resummation at next-to-next-to-leading logarithmic
  accuracy}},  {\em Phys.Lett.} {\bf B696} (2011) 207--213,
  [\href{http://xxx.lanl.gov/abs/1007.2351}{{\tt arXiv:1007.2351}}].

\bibitem{Harlander:2014hya}
R.~V. Harlander, A.~Tripathi, and M.~Wiesemann, {\it {Higgs production in
  bottom quark annihilation: Transverse momentum distribution at NNLO$+$NNLL}},
   {\em Phys.Rev.} {\bf D90} (2014), no.~1 015017,
  [\href{http://xxx.lanl.gov/abs/1403.7196}{{\tt arXiv:1403.7196}}].

\bibitem{Harlander:2014uea}
R.~V. Harlander, H.~Mantler, and M.~Wiesemann, {\it {Transverse momentum
  resummation for Higgs production via gluon fusion in the MSSM}},  {\em JHEP}
  {\bf 1411} (2014) 116, [\href{http://xxx.lanl.gov/abs/1409.0531}{{\tt
  arXiv:1409.0531}}].

\bibitem{Monni:2014zra}
P.~F. Monni and G.~Zanderighi, {\it {On the excess in the inclusive
  $W^+W^-\rightarrow l^+l^-\nu\bar\nu$ cross section}},  {\em JHEP} {\bf 1505}
  (2015) 013, [\href{http://xxx.lanl.gov/abs/1410.4745}{{\tt
  arXiv:1410.4745}}].

\bibitem{Jaiswal:2014yba}
P.~Jaiswal and T.~Okui, {\it {Explanation of the $WW$ excess at the LHC by
  jet-veto resummation}},  {\em Phys.Rev.} {\bf D90} (2014), no.~7 073009,
  [\href{http://xxx.lanl.gov/abs/1407.4537}{{\tt arXiv:1407.4537}}].

\bibitem{Becher:2014aya}
T.~Becher, R.~Frederix, M.~Neubert, and L.~Rothen, {\it {Automated NNLL $+$ NLO
  resummation for jet-veto cross sections}},  {\em Eur.Phys.J.} {\bf C75}
  (2015), no.~4 154, [\href{http://xxx.lanl.gov/abs/1412.8408}{{\tt
  arXiv:1412.8408}}].

\bibitem{Nason:2004rx}
P.~Nason, {\it {A New method for combining NLO QCD with shower Monte Carlo
  algorithms}},  {\em JHEP} {\bf 0411} (2004) 040,
  [\href{http://xxx.lanl.gov/abs/hep-ph/0409146}{{\tt hep-ph/0409146}}].

\bibitem{Sjostrand:2006za}
T.~Sjostrand, S.~Mrenna, and P.~Z. Skands, {\it {PYTHIA 6.4 Physics and
  Manual}},  {\em JHEP} {\bf 0605} (2006) 026,
  [\href{http://xxx.lanl.gov/abs/hep-ph/0603175}{{\tt hep-ph/0603175}}].

\end{thebibliography}\endgroup

\end{document}